\documentclass[%
 reprint, 
 groupedaddress,
 amsmath,amssymb,
 aps,
 pre,
]{revtex4-1}

\usepackage{graphicx}
\usepackage{dcolumn}
\usepackage{bm}
\usepackage{hyperref}
\usepackage[mathlines]{lineno}

\newcommand{\sci}[2]{#1\times 10^{#2}}
\newcommand{\abs}[1]{\left|#1\right|}
\newcommand{\const}{\mathit{const}}

\DeclareMathOperator{\Real}{Re}
\newcommand{\xunit}{\hat{x}}
\newcommand{\rhosurftip}{\rho_s}
\newcommand{\aph}{a_\mathrm{ph}}
\newcommand{\Feff}{F_{\aph}}
\newcommand{\rvec}{\mathbf{r}}
\newcommand{\Evec}{\mathbf{E}}
\newcommand{\Jvec}{\mathbf{J}}
\newcommand{\vvec}{\mathbf{v}}
\newcommand{\excess}{\mathrm{exc}}
\newcommand{\attach}{\mathrm{att}}

\begin{document}


\title{Electric streamers as a nonlinear instability: the model details}


\author{Nikolai G. Lehtinen}
\email[]{Nikolai.Lehtinen@uib.no}
\affiliation{Birkeland Centre for Space Science, University of Bergen, Bergen, Norway}


\date{\today}

\begin{abstract}
We propose a new approach to unambiguous determination of parameters of positive and negative electric streamer discharges. From hydrodynamic equations, in the assumption of a solution in the shape of a streamer, it is possible to derive several relations between streamer parameters, which form a system of algebraic equations (SAE). Because of the made approximations, the error in the solution of this system is expected to be probably up to a few tens of percent. Solving the SAE allows us to express all streamer parameters in terms of the streamer length $L$, the constant uniform external electric field $E_e$, and the streamer radius. The solutions with different radii are valid solutions of the hydrodynamic equations, and are analogous to the propagation modes of flat-front perturbations with different transverse wavelengths. We interpret the streamer as a nonlinear instability, whose behavior is determined by choosing the radius at which the velocity is maximized, because, as we show, the velocity plays the same role as the exponential growth rate in the case of linear instabilities.

Thus, streamer behavior is unambiguously determined by $E_e$ and $L$, in a relatively computationally economical way. In contrast, numerical methods of solving the microscopic equations, such as hydrodynamic simulations, are more computationally demanding, and the preferred solution in them arises automatically because of numerical fluctuations. The calculations for air at sea level conditions produce reasonable values for commonly observed streamer parameters. The calculated positive streamer velocities and negative threshold fields are compatible with experimental measurements. The physical reason for the positive threshold fields is also discussed. A much simplified analytical model (Appendix~\ref{app:analytic}) reproduces many of the presented results, at least qualitatively.
\end{abstract}

\pacs{}

\maketitle

\section{Introduction}
\label{sec:introduction}

\subsection{What determines the streamer radius?}

Electric streamer discharges are ionized columns in gas or liquid which advance by ionizing the material in front of them with the enhanced field at the streamer tip \citep{book:Bazelyan+Raizer1998,book:Raizer1991}. They are an important stage in the formation of sparks, and thus, especially those propagating in air, play a huge role both in technology and natural phenomena.

The physics determining the parameters of a streamer discharge in air, such as its radius and speed, has been a long-standing problem \citep{Ebert+Sentman2008}. As \citet[p.~46--47]{book:Bazelyan+Raizer1998}, formulated it, ``The mechanisms by which a plasma conductor acquires a definite [...] radius [...] seem to go far beyond the steady state processes [...] We should recognize that these mechanisms are not quite clear at present.''  Uncovering these mechanisms is the goal of the present paper.

\citet{book:Loeb+Meek1941} were the first to propose that electrons undergoing impact ionization avalanche in high electric field in air create sufficient space charge to form a streamer. They also suggested that the initial size of the streamer was determined by the transverse spreading of the electrons in the avalanche due to diffusion. This idea was taken up by other researchers, who used the spreading due to diffusion to estimate the streamer radius not only at its formation, but also during its propagation \citep{Dawson+Winn1965, Gallimberti1972, Qin+Pasko2014}. One may estimate, however, that spreading of the streamer due to diffusion is much slower than that suggested by observations and computer simulations.  The transverse size of a flux of directed velocity $V$ diffusing with coefficient $D$ grows as $\sqrt{D L/V}$ with distance $L$. Substituting typical values for a laboratory streamer in air, $L\sim 0.1$~m, $V\sim 10^6$~m/s and $D\sim 0.1$~m$^2$/s, we get the transverse size increase of $\lesssim$0.1~mm while typical observations show radii $>$1~mm \citep[e.g.,][]{Chen+2013,Yi+Williams2002,Briels+2008}. Moreover, \citet{Naidis2009} argued that diffusion may be completely neglected in the approximate analysis of regular streamer propagation. Thus, diffusion is probably not the right explanation of the streamer radius. The argument that the electrostatic repulsion of electrons in the highly ionized streamer head leads to the increase of the radius is also not valid, even in the negative streamer case, because the displaced electrons leave behind positive ion charge that pulls them back.

Streamer parameters, such as its speed and the transverse size (radius), may be determined in a numerical experiment by solving microscopic physics equations, assuming that the methods used are stable and accurate. Examples of computationally-intensive numerical approaches include, e.g., adaptive mesh refinement 3D hydrodynamic models \citep{Teunissen+Ebert2017} and PIC (Particle-In-Cell) simulations \citep{Chanrion+Neubert2008}. More information can be found in extensive reviews \citep{Ebert+Sentman2008,Ebert+2010,Pasko+2013,Raja+2018,book:Pasko2006}. Numerical finite-difference hydrodynamic streamer models considered by \citet{Bagheri+2018} were in code verification (i.e., internal consistency check) stage, but not in validation (i.e., quantitative agreement with experiment) stage. Depending on setup, they showed considerable variations ($\sim 10$\% inferred from figures in \citep{Bagheri+2018}) due to numerical errors and were plagued by numerical instabilities, appearing as oscillations. Besides, even though such simulations reproduce the correct columnar streamer shapes and the order of magnitude of experimentally measured streamer parameters, they still leave open the question of what physical principles determine them.

Another approach to determine the typical transverse size is the perturbative analysis of ionization fronts. The analysis of flat ionization fronts yielded many useful results, such as constraints on the possible values of the front speed and the relation between the field ahead of the front and the ionization behind it \citep{book:Lagarkov+Rutkevich1994}. Different-size transverse harmonic perturbations of such a front (modes) may grow exponentially at different rates. There have been multiple studies with intention to relate the transverse size of a streamer to the size of the fastest growing transverse perturbation of a flat or a curved front \citep{Arrayas+Ebert2004,Derks+2008,Arrayas+Fontelos2011}. The most comprehensive of these is the work by \citet{Derks+2008}, who considered hydrodynamic equations describing a flat ionization front with small transverse harmonic perturbations, including both electron drift and diffusion, and calculated growth rates as a function of the transverse wavenumber (or, equivalently, transverse wavelength). The preferred transverse size (i.e., the one at which the growth of an instability is maximized) was calculated to be $\propto D^{1/4}$ \citep{Ebert+Derks2008}. Unfortunately, there has been no continuation of flat-front perturbation studies (to our knowledge) to include photoionization, which is a nonlocal effect and therefore is much harder to tackle than diffusion.

\subsection{Overview of the used method}

Our approach is somewhat analogous to the flat-front analysis of \citet{Derks+2008}. We also start with a system of hydrodynamic PDE, including all the relevant physics. Unlike \citet{Derks+2008}, we do include photoionization, because it is crucial for streamers in air, because we are aiming to obtain practical results for air discharges. As in the flat-front case, we also look for a solution in a particular geometric shape, but instead of a harmonic shape of a small flat-front perturbation, we look for a solution in the shape of a streamer, i.e., a cylindrical column. The system of PDE, by making approximations, is eventually reduced to a finite system of algebraic equations (SAE) with a finite set of unknowns, which include simple measurable streamer parameters, such as velocity and radius. As in the flat-front case, where the transverse wavelength was arbitrary, an unambiguous answer cannot be obtained, but we get a set of streamer ``modes'' corresponding to different streamer radii $a$. Each such mode is a valid solution with its own set of parameters, e.g. streamer speed $V(a)$. An extra criterion is thus needed for selecting the ``real'' set of parameters. In a flat-front theory \citep{Derks+2008} (or, in fact, in any linear unstable system), there is a preferred solution, characterized by the maximum growth rate. It arises from initial conditions with arbitrary random small fluctuations, which are present because the initial conditions cannot be specified with infinite precision. In Section~\ref{sec:maxv} we argue that, in our case, the preferred streamer mode is found by maximizing $V(a)$.

An approach with reduction of microscopic equations into a SAE had been attempted previously by other authors \citep[e.g.][]{Dyakonov+Kachorovskii1989}, but usually was met with readers' frustration. This was expressed, e.g., by \citet[Preface]{book:Bazelyan+Raizer1998}: in their experience, many ``readers ... admire formulas without understanding their physical meaning, but more experienced researchers would rather start thinking of a theory of their own.'' The previous SAE formulations are discussed and criticized in Subsection~\ref{ssec:other_authors}. When finding the streamer modes, we implement the program of \citet[p.~46--47]{book:Bazelyan+Raizer1998} who proposed to separate all physics that determine streamer propagation into two problems: what happens at the streamer tip and what is the role of the streamer channel. We solve these two problems simultaneously. We will find that the channel affects the processes at tip by the values of its length and intrinsic field, while the parameters at the tip are related to each other by other relations, all of which will be collected in SAE (\ref{eq:system}). Approximations used in deriving the simplified equations are stated in Subsection~\ref{ssec:general}, and additional approximations are introduced and discussed throughout the text. Our model is numerical; however, most qualitative results may be demonstrated with a simpler (but more approximate) analytical solution, presented in Appendix~\ref{app:analytic}. Thus, in this work, we aim at creating the least controversial theory of streamer parameters that is based on well-known and established equations and principles.

\section{The streamer model}
\label{sec:model}

\subsection{Overview and the most important approximations and notations}
\label{ssec:general}

\subsubsection{Discussion of hydrodynamic approximation}
\label{sssec:nonlocality}
The most accurate description of an electric discharge in air would be provided by solving equations of motion of all particles and electromagnetic fields. Of course, this task is impossible and certain approximations must be made. We use electrostatic approximation (i.e., electric field $\Evec=-\nabla\phi$ is a potential field) because the expected maximum velocities of streamers ($\sim 10^6$~m/s) are well below the speed of light. For motion of electrons, with very good accuracy one can use hydrodynamic equations, which were solved numerically for a realistic streamer first (even though without photoionization) by \citet{Dhali+Williams1987}. Such hydrodynamic equations neglect nonlocal effects due to electron transport, a more accurate description of which would be provided by a kinetic equation, or by numerically following individual particles as in PIC methods.  The nonlocal effects may be crucial in extreme conditions of streamer propagation, e.g. during the avalanche-streamer transition or streamer collision process \citep{Lehtinen+Ostgaard2018}, but only provide a correction in regular conditions of streamer propagation \citep{Naidis1997}. For example, Figure~3 of \citet{Dujko+2011} shows that the percentage difference between the bulk and flux components of the drift velocity may be as high as 20\% at $E\sim 10$~MV/m in sea-level air, which is easily achievable in the head of a streamer. The correction due to nonlocal effects may be as high as tens of percent for the electric field in front of the streamer, as calculated by \citet[Figure~1a]{Naidis1997}, and is probably due to the steepening of the ionization front. Other physical variables are affected to a lesser degree \citep[Figures~1b,~2]{Naidis1997}. Such accuracy, however, is beyond the goal of the present paper.

\begin{figure}
\includegraphics[width=0.49\textwidth]{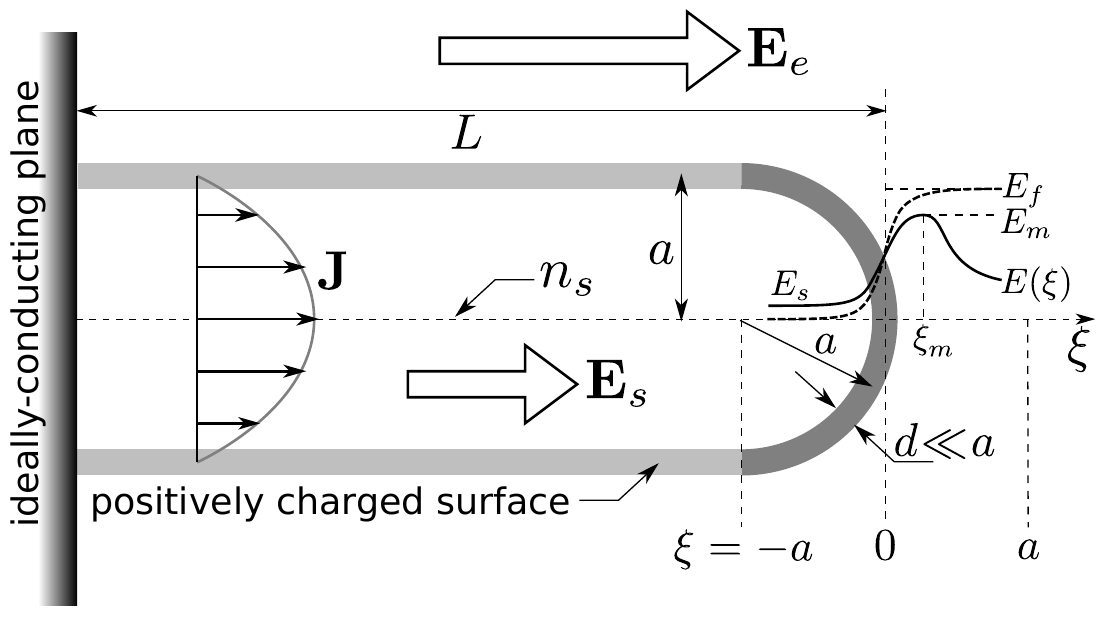}
\caption{The streamer model. The field directions are for a positive streamer (for a negative streamer they are opposite).
\label{fig:sheath}}
\end{figure}

\subsubsection{Assumed geometric shape and basic streamer parameters}
We look for a solution of hydrodynamic equations for an ionization front in the shape of a propagating (growing) ionized column (a \emph{streamer}), sketched in Figure~\ref{fig:sheath}. The streamer is immersed in given external constant uniform field $\Evec_e\parallel\xunit$, , where $\xunit$ is the unit vector along $x$-axis, and has a cylindrically symmetric shape with the axis of the streamer $\parallel\xunit$. That streamers have such a shape, or close to it, is well known from experiments and numerical simulations. The streamer is attached to an ideally conducting plane $\perp\xunit$. The plane absorbs electrons from the streamer in the case of a positive (cathode-directed) streamer, and is assumed to be an ideal emitter of electrons in the area covered by the channel in the case of a negative (anode-directed) streamer. The interaction with the electrode in the case of a negative streamer and a non-ideal electron emitter was not considered in this paper and is a subject of future research. The streamer \emph{head} is a hemisphere of radius $a$ attached at the \emph{neck} to the \emph{channel}, which is represented as a long cylinder of the same radius. Possible variations of the channel radius along its length as a source of error in the presented model are discussed in Subsubsection~\ref{sssec:a_variation} and in Subsubsection~\ref{sssec:Es_ns_uniformity}. The total length (including the head) is denoted $L$. By construction it is necessary that $L>a$, and because streamers are usually rather narrow and long columns, we will use approximation $L\gg a$ when necessary. The interior carries a uniform \emph{intrinsic field} $\Evec_s$, which is lower than the external field. Electron density $n$ is $>0$ inside the streamer and quickly drops (with a scale $d\ll a$) to $n=0$ outside. Inside the channel, $n$ is assumed to be uniform on the axis and its value is denoted $n_s$. Electron density decreases towards the channel walls; in Figure~\ref{fig:sheath}, the transverse profile of $n$ is proportional to $\Jvec$, which is drawn as approximately parabolic. Uniformity of $\Evec_s$ and $n_s$ is an approximation, deviations from which are discussed in Subsubsection~\ref{sssec:Es_ns_uniformity}. The spatial charges are concentrated mostly on the surface of the streamer. We will demonstrate in Subsubsection~\ref{sssec:relaxation} that the bulk charge relaxation time due to conductivity inside the channel (the Maxwellian time scale) is very short, so it is valid to assume spatial charge neutrality of the interior of the channel. The system of the depicted streamer and the ideally-conducting plane is equivalent electrostatically to a conducting rod of length $2L$ with hemispherical caps on both ends, suspended in free space, i.e., the original streamer plus its electric image in the plane.

The streamer grows (propagates) along $x$-axis with velocity $V\equiv d L/d t$. The propagation is aligned with $\Evec_e$ in the case of a positive streamer, and anti-aligned in the case of a negative streamer. The characteristics of the head (radius, fields around and inside it, etc.) change relatively slowly  (i.e., $d a/d t\ll V$ if $L\gg a$, etc.) so we will assume that the head is in a stationary state in its moving reference frame. We will use co-moving coordinate $\xi=x - V t$, so that for stationary propagation $\partial_x=\partial_\xi$ and $\partial_t = -V \partial_\xi$, where $\partial$ denotes the derivative in respect to the variable indicated by the subscript. We will not use $x$ coordinate very often so we do not specify its origin, but we fix $\xi=0$ to correspond to the \emph{streamer tip}, which is the foremost part of the \emph{streamer front} (the curved ionization front, located at the surface of the hemisphere corresponding to the head). \emph{Front thickness} $d$ is defined as the typical scale ($e$-folding distance) on which electron density $n(\xi)$ drops from $n=n_s$ inside the channel to zero outside the streamer, and is much smaller than the streamer radius, $d\ll a$. We will only consider values of electric field on the axis of the streamer. On the axis, due to cylindrical symmetry, all fields are $\parallel\xunit$. This allows us to introduce scalar notation $E$ denoting the signed $x$-component of the field instead of working with vector $\Evec$ (and analogously for other vectors). Namely, we denote $E=\pm E_x$  (or, equivalently, $\Evec=\pm E\xunit$) on the axis of a positive (negative) streamer so that always $E>0$. (Also, $\Evec_s=\pm E_s\xunit$ with $E_s>0$ inside the channel.) Throughout this paper, when we use $\pm$ or $\mp$, the upper sign will correspond to a positive, and lower to a negative streamer. Neither $E$ nor spatial charge ever reverse their sign, as indicated by the results of hydrodynamic simulations \citep[e.g.][]{Lehtinen+Ostgaard2018}. (The underlying explanation may be that a field or charge density reversal would raise the energy of electric field, which is only possible if electrons can pump their kinetic energy into it. However, this is impossible in the hydrodynamic approximation that we use, because electron motion is non-inertial.) This means, e.g., that a positive streamer only has positive spatial charges associated with it.

The total error associated with the simplifying assumptions made here is probably of the order of a few tens of percent (as suggested in Subsubsection~\ref{sssec:Es_ns_uniformity}).

\subsubsection{Microscopic processes and hydrodynamic equations}
We consider the following transport and reaction processes. Most of electron density growth in the streamer is due to avalanche impact ionization in the streamer front, described by temporal ionization rate $\nu_i(E)$. The electrons attach to neutrals, forming negative ions, with temporal rate $\nu_a(E)$. The total electron production (\emph{net ionization}) rate is
\begin{equation}
 \nu_t\equiv \nu_i-\nu_a
 \label{eq:nu_t}
\end{equation}
The seeds for the impact ionization avalanches are provided by the photoionization process with source $s_p$, which is described in Subsection~\ref{ssec:photo}. The streamer propagation is affected by electron drift, velocity of which is $\mathbf{v} = -\mu(E) \mathbf{E}$, where $\mu(E)>0$ is the electron mobility. In scalar notations for values on axis, $v=\mu(E) E$ with $v>0$, so that $\mathbf{v}=\mp v\hat{x}$. The values of coefficients $\nu_i$, $\nu_a$ and $\mu$ used in calculations are given in Subsection~\ref{ssec:air_coefs}. The electron drift is also responsible for the electrical conductivity, since the ion mobility is about $\sim$1000 times lower. The conductivity current density is given by $\Jvec_c=-e\vvec n$ or $J_c=e v n$ ($\Jvec_c=\pm J_c\hat{x}$ on axis), where $e>0$ is the absolute value of the electron charge. In a system with changing field, the displacement current $\Jvec_d = \varepsilon_0 \partial_t \Evec$ ($\Jvec_d=\pm J_d\hat{x}$ on axis) is also important, so we reserve the notation $\Jvec = \Jvec_c + \Jvec_d$ for the total current ($\Jvec=\pm J\hat{x}$ on axis), which is divergenceless as a general result of Maxwell's equations:
\begin{equation}
  \nabla\cdot\Jvec=0,\quad\Jvec = \Jvec_d+\Jvec_c = \varepsilon_0\partial_t\Evec - e\vvec n
  \label{eq:Jdiv0}
\end{equation}

We neglect electron diffusion, as it is not important for later stages of streamer propagation \citep{Naidis2009}, as well as nonlocality in electron transport \citep{Naidis1997}, which were discussed in Subsubsection~\ref{sssec:nonlocality}.

The hydrodynamic equations for particle densities are \citep[e.g.,][]{Morrow+Lowke1997}:
\begin{equation}
  \left.
  \begin{array}{rcl}
    \varepsilon_0\nabla\cdot\Evec & = & e(n_i-n) \\
    \partial_t n + \nabla\cdot(\vvec n) & = & \nu_t n + s_p \\
    \partial_t n_i & = & \nu_t n + s_p
  \end{array}\right\}
  \label{eq:hydro}
\end{equation}
where we introduce the net ion density $n_i$ (positive minus negative), needed for charge balance. We neglect electron detachment from negative ions, recombination and inter-ion processes. (These processes may, however, be important in the streamer channel, see discussion in Section~\ref{ssec:att}.)

We can exclude $n_i$ from the system (\ref{eq:hydro}) by subtracting the last two equations and substituting the result into the first, obtaining
\begin{equation}
  \left.
  \begin{array}{rcl}
    \varepsilon_0\partial_t\Evec &=& e\vvec n + \Jvec \\
    \partial_t n + \nabla\cdot(\vvec n) & = & \nu_t n + s_p
  \end{array}\right\}
  \label{eq:hydro_front}
\end{equation}
which is more convenient for solving if $\Jvec$ is somehow specified.

\subsubsection{Additional streamer parameters}
There are three more notations that are used throughout this paper and thus for convenience are listed here, even though they are introduced later:
\begin{enumerate}
\item \emph{Maximum field} at the streamer tip $E_m$, see Figure~\ref{fig:sheath};
\item \emph{Idealized maximum field} or \emph{flat-front maximum field} in front of the streamer $E_f$ which would be obtained for $d\rightarrow 0$ or if the front were completely flat ($a=\infty$). Due to finite thickness $d>0$ and curvature of the streamer head ($a<\infty$), however, we have $E_m<E_f$. The relation between $E_f$ and $E_m$ is obtained in Appendix~\ref{app:ns} and is given by equation (\ref{eq:Em}). Both the real field and the flat-front field are sketched in Figure~\ref{fig:sheath} with solid and dashed lines, respectively. Field $E_f$ is a measure of the surface charge density at the streamer tip, $\rhosurftip=\varepsilon_0 (E_f-E_s)$, and is calculated in Subsubsection~\ref{sssec:Ef}.
\item The \emph{electric field width} at the tip of the streamer $l$, which is of the order of a fraction of $a$. It was defined previously, e.g., by \citet{Naidis2009} and is introduced in Subsubsection~\ref{sssec:Exi}.
\end{enumerate}

\subsection{The $E$ field outside the ionization front}

\subsubsection{Field enhancement at the tip}
\label{sssec:Ef}
The electric field for $\abs{\xi}\gg d$ is calculated in the assumption of an infinitely thin surface ($d=0$) using the cylindrically-symmetric implementation of the method of moments (MoM) \citep[ch.~2]{book:Harrington1993} which was developed by the author for the paper of \citet{Skeltved+2017} where its description may be found. The charges are redistributed on the surface of the streamer in order to create uniform field $\Evec_s$ inside it. Finding the surface charges in such a configuration is equivalent to finding them for an ideally-conducting rod (with zero field inside) if we subtract $\Evec_s$ everywhere. The resulting excess field is $\Evec^\excess\equiv \Evec-\Evec_s$. The excess external field $E_e^\excess=E_e-E_s$ is then enhanced by the charges on the streamer surface near the streamer tip, and canceled by them inside the streamer. We denote the ratio of the field created (only) by these charges just at the streamer tip ($\xi=+0$) to the excess external field as the \emph{field enhancement factor}, $\eta\equiv\eta(L/a)$, which is a function of geometry only, i.e. of the ratio of the streamer length to its radius. Thus, by definition $E^\excess_\mathrm{charges}(+0)=E^\excess(+0)-E_e^\excess=E_e^\excess\eta$. With notation $E_f\equiv E(+0)$, this becomes
\begin{equation}
  E_f = E_s + (E_e-E_s) (\eta + 1) = E_e + (E_e-E_s) \eta
  \label{eq:Ef}
\end{equation}

The MoM results for $\eta$ are close (within 4\% for $L/a>5$) to those given by an approximate formula of \citet[p.~78]{book:Bazelyan+Raizer1998}:
\begin{equation}
  \eta(L/a) \approx 2+0.56(2L/a)^{0.92}
  \label{eq:eta}
\end{equation}

\subsubsection{Field outside the tip}
\label{sssec:Exi}
The calculated dependence on $\xi$ of the electric field created by the streamer charges is fitted to the shape $\propto 1/(\xi+l)$ proposed by \citet{Naidis2009} in region $0<\xi<a$. We will demonstrate in Subsection~\ref{ssec:photodistance} that the region $\xi>a$ is not important for streamer development, so we can write
\begin{equation}
  E(\xi>0)|_{d=0} \approx E_e + \frac{(E_e-E_s)\eta}{1+\xi/l} = E_e + \frac{E_f-E_e}{1+\xi/l}
  \label{eq:E_approx}
\end{equation}
Taking into account smooth transition from inside to outside streamer with finite front thickness $d>0$, we can write this as
\begin{equation}
  E(\xi) = E_s + (E_e-E_s)S(\xi)\left[\frac{\eta}{1+\xi/l} + 1\right]
  \label{eq:E}
\end{equation}
where the ``switch function'' $S(\xi)$ is such that $S(\xi\ll-d)=0$ and $S(\xi\gg d)=1$ and describes the shape of the front of thickness $d$. The effects of finite $d$ are discussed in detail in Subsection~\ref{ssec:front} and Appendix~\ref{app:ns}. For $d\rightarrow 0$, the switch function becomes the Heaviside step function $S(\xi)=\Omega(\xi)$, defined by $\Omega(\xi>0)=1$ and $\Omega(\xi<0)=0$.

Far ahead of the front, $S(\xi)\approx 1$, so that we can use formula (\ref{eq:E_approx}). This is valid at $\xi\gtrsim\xi_m$, where $\xi_m$ is the location where the field is maximal (see Figure~\ref{fig:sheath}) and is such that $d<\xi_m<l$. At this point, $\xi_m$ still remains to be found; but looking ahead, it is defined in equation (\ref{eq:Em}). Thus,
\begin{equation}
  \left.E(\xi\gtrsim\xi_m)\right|_{d>0} = \mbox{equation (\ref{eq:E_approx})}
  \label{eq:E_approx_1}
\end{equation}

The parameter $l/a$ (taken as $\approx$0.33--0.5 by \citet{Naidis2009}) was calculated with MoM so that it gives the best fit for equation (\ref{eq:E_approx}) at $0<\xi<a$, and was fitted with the formula
\begin{equation}
  l/a \approx 0.40 - \frac{0.59}{(L/a) + 2.31}
  \label{eq:l}
\end{equation}

\subsubsection{Linear charge density}
Another important output of the MoM is the linear charge density (per unit $\xi$) along the streamer channel $\lambda(\xi)$, which equals to the surface charge density on the surface of the channel integrated over the azimuth. The linear charge density is variable, growing from zero at the beginning (i.e., tail) of the streamer (defined as the location where it is attached to the conducting plane, i.e. $\xi=-L$) to the maximum at the streamer tip. The value of $\lambda(\xi)$ is at the streamer neck, $\lambda_\mathrm{neck}\equiv\lambda(\xi=-a)$, was calculated numerically with MoM and approximated with a fit
\begin{equation}
  \lambda_\mathrm{neck}\approx \varepsilon_0 (E_e-E_s) a \left[2.70\left(L/a-1\right) + 7.43\sqrt{L/a-1}\right]
  \label{eq:lambda_neck}
\end{equation}

The field of the shape given by (\ref{eq:E_approx}) can be created by a uniform linear charge density concentrated on the semi-axis at $\xi<-l$ with value
\begin{equation}
  \lambda_E = 4\pi \varepsilon_0 l (E_f-E_e) = 4\pi \varepsilon_0 l (E_e-E_s) \eta
  \label{eq:lambda_E}
\end{equation}
which is actually very close (within 7.5\% error) to $\lambda_\mathrm{neck}$ given by (\ref{eq:lambda_neck}) for all values of $L/a$. Even though the real linear charge density is nonconstant, it varies with a large scale $L\gg a$, so this variation is not important when we consider fields close to the tip, $\xi\lesssim a$.

\subsubsection{Uniformity of $a$ along the channel and the field near the tip}
\label{sssec:a_variation}
We assumed that the streamer channel had constant radius throughout its length. If $E_s=\const$, the variations of the channel radius far from the streamer head are not going to affect much the field outside the tip, $E(\xi>0)$, because it is formed mostly by the surface charges near the tip (e.g., uniform linear density in equation (\ref{eq:lambda_E})). The uniformity of $E_s$ is discussed in Subsubsection~\ref{sssec:Es_ns_uniformity}.

\subsection{Electrical currents}

\subsubsection{Current continuity on axis}
Let us look qualitatively at the structure of the total current density $\Jvec = \Jvec_c + \Jvec_d$, where $\Jvec_c = -e\vvec n$ is the conductivity current and $\Jvec_d = \varepsilon_0 \partial_t \Evec$ is the displacement current. The current inside the streamer channel is mostly the conductivity current which has the highest value of $J_c\approx e n_s v(E_s)$ on the streamer axis and may be distributed non-uniformly in the radial direction; the displacement current is small because of approximate uniformity of $\Evec_s$. The current outside the streamer, i.e., at $\xi>\xi_m$, where equation (\ref{eq:E_approx}) is also valid according to equation (\ref{eq:E_approx_1}), is mostly the displacement current due to absence of free electrons in that region:
\[ J(\xi) = \varepsilon_0 \partial_t E = -\varepsilon_0 V \partial_\xi E \]
Substituting (\ref{eq:E_approx}), and integrating, we get
\[ J(\xi) = \frac{J_0}{(1+\xi/l)^2},\quad\xi>\xi_m \]
where
\begin{equation}
   J_0 = \frac{\varepsilon_0 V (E_e-E_s)\eta}{l} = \frac{\varepsilon_0 V (E_f-E_e)}{l}
  \label{eq:J0}
\end{equation}
This is a current that diverges in three dimensions, and it is natural to assume that it starts diverging from the tip of the streamer, i.e., the above expression $\propto (1+\xi/l)^{-2}$ is approximately valid also for $0<\xi<\xi_m$. Inside the streamer ($\xi<0$), the current is the conductivity current and is approximately constant, and taking into account continuity at $\xi=0$, we get
\begin{equation}
  J(\xi)=J_0\times\left\{\begin{array}{lr} (1+\xi/l)^{-2}, & \xi>0 \\ 1, & \xi<0 \end{array}\right.
  \label{eq:J}
\end{equation}
with $J_0$ given by (\ref{eq:J0}). The conductivity current is $J_c = e n_s v(E_s)$, so the current continuity relation at $\xi=0$ becomes
\begin{equation}
  e n_s v(E_s) = J_0 = \frac{\varepsilon_0 V (E_e-E_s)\eta}{l}
  \label{eq:Jcontinuity}
\end{equation}
This is essentially the same as equation~(6) of \citet{Babaeva+Naidis1997} if we use the second expression for $J_0$ from (\ref{eq:J0}).

\subsubsection{Discussion of conductivity and displacement current variations inside the front}
Although the total current density is continuous (and $\approx J_0$ on axis) through the thin ionization front ($\abs{\xi}\lesssim d$), both conductivity and displacement current, when taken separately, greatly exceed the total value $J_0$ in that region, but almost cancel each other. For a positive streamer, the conductivity current is aligned in $\xi$ direction and is due to the very high field $\sim E_f$ and high electron density $\sim n_s$, while the displacement current is in the direction opposite to $\xi$ and is due to the sudden drop of electric field from $E_f$ to $E_s$. For a negative streamer, the directions are reversed.

\subsubsection{Current integrated over the cross-sectional area}
\label{sssec:currents_integrated}
The total channel current, integrated over the cross-sectional area of the channel, is $I = \pi a_I^2 J_c$, where $a_I$ is the effective radius $<a$. We can approximately determine the value of $a_I$ from equation (\ref{eq:Jcontinuity}). The total channel current $I$ is the same as given by the transportation of the charged channel with the velocity of the streamer, i.e., $I=\lambda V$. This must hold at least at the neck, with $\lambda=\lambda_\mathrm{neck}$ given by equation (\ref{eq:lambda_neck}), or we can use the uniform linear charge density given by equation (\ref{eq:lambda_E}) so that $I=\lambda_E V=\const$. Equating this to the conductivity current, and using (\ref{eq:Jcontinuity}), we get
\[  I = \pi a_I^2 e n_s v(E_s) = \pi a_I^2 \frac{\varepsilon_0 V (E_e-E_s)\eta}{l} \approx \lambda_E V \]
Thus,
\[ \frac{a_I}{l} \approx \sqrt{\frac{\lambda_E}{\pi\varepsilon_0l(E_e-E_s)\eta}} = 2 \]
where we used (\ref{eq:lambda_E}).

The simulations \citep[e.g.][]{Lehtinen+Ostgaard2018} indicate that the radial current distribution is approximately parabolic (as sketched in Figure~\ref{fig:sheath}), which would give $a_I\approx a/\sqrt{2}$. This would be obtained for
\begin{equation}
  l/a\approx 1/\sqrt{8}
  \label{eq:l_approx}
\end{equation}
which matches equation (\ref{eq:l}) well for a wide range of $L/a$.

We may perform a sanity check to see if the total current flowing in the streamer channel is conserved. The displacement current flows out approximately isotropically. (A uniformly charged thin moving rod creates an isotropic displacement current flowing out of its end.) The center from which the isotropic displacement current is flowing may be taken at $\xi=-l$, as suggested by equation (\ref{eq:J}).  The total current is integrated over the surface of a sphere of radius $\xi+l$ centered at $\xi=-l$ and is equal to $I=4\pi(\xi+l)^2 J(\xi) = 4\pi J_0 l^2$ which matches $I=\lambda V$ if $\lambda=\lambda_E$ is given by equation (\ref{eq:lambda_E}).

\subsubsection{On the assumptions of uniformity of $E_s$ and $n_s$ along the channel}
\label{sssec:Es_ns_uniformity}
In order to calculate the electric field outside the streamer tip in Subsubsection~\ref{sssec:Exi}, we assumed $E_s=\const$. Equation (\ref{eq:Jcontinuity}) requires constant current along the channel, i.e., $n_s v(E_s)=\const$, which together with $E_s=\const$ means that we also need $n_s=\const$. However, the assumed uniformity of $E_s$ and $n_s$ along the streamer channel may be invalidated by the neglected attachment and recombination processes, which is the subject of Subsection~\ref{ssec:att}. Alternatively, it could be affected by (1)~large variations of the channel radius far from the streamer head; or (2)~the history of channel development. Let us discuss both of these possible causes.

\begin{enumerate}
\item The variation of channel radius along the channel may affect the conductivity current flowing through the channel. In particular, both observations and hydrodynamic simulations show that the channel widens as the the streamer propagates, thus having a conical shape which narrows towards the streamer tail. However, it is still possible to have both the total current $I\approx\const$ (see Subsubsection~\ref{sssec:currents_integrated}) and $E_s\approx\const$ even in this situation if somehow the integral of electron density $n$ over the transverse area of the channel is kept constant, and electrons flow through the variable-radius channel like through a pipe, keeping the same drift velocity determined by $E_s$.
\item The tail part of the channel was created earlier and at different conditions, and therefore $n_s$ there may be different from that in the parts closer to the head which were created later. (We may assume that $n_s$ locally stays the same because the electrons in the channel do not drift much on the time scale of the streamer development, i.e. $v(E_s)\ll V$.) The evolution of $n_s$ may be inferred if we look ahead, at Figures~\ref{fig:pos_results}e and~\ref{fig:neg_results}e, from which we see that $n_s$ may either decrease (for positive streamers) or grow (for negative streamers) with streamer length. A non-stationary calculation is beyond the scope of the presented paper. The error associated with this approximation comes from replacing the average $n_s$ along the channel with the value from latest added streamer segment, and may be up to a few tens of percent. This may be the biggest source of error in the presented model.
\end{enumerate}

\subsection{$E$ inside the ionization front}
\label{ssec:front}
The value $E_f$ given by equation (\ref{eq:Ef}) is actually never achieved due to finite size of front thickness $d$. In fact, the maximum field $E_m=E(\xi_m)<E_f$ is achieved at $\xi_m$, such that $d<\xi_m<a$ (see the sketch of the plot of $E(\xi)$ in Figure~\ref{fig:sheath}). However, field $E_f$ still has a physical meaning: together with $E_s$ it determines the surface charge density at the tip of the streamer $\rhosurftip = \int \rho\,d\xi = \varepsilon_0(E_f-E_s) = \varepsilon_0(E_e-E_s)(\eta+1)$, where the integration is through the ionization front. Thus, it would be the correct field ahead of the front, if the front were exactly flat. (The sketch of the flat-front field is plotted with a dashed line in Figure~\ref{fig:sheath}.)

\subsubsection{Relation between $n_s$ and $E_f$ for the flat front}
\label{sssec:ns}
The flat ionization front had been studied in great detail \citep[see, e.g.][ch. 3]{book:Lagarkov+Rutkevich1994}. In this Section, we quickly review its theory, in order to make the connection between $n_s$ and $E_f$. In a flat front, the total current $\nabla\cdot\Jvec=\partial_\xi J = 0$ so necessarily $J=\const=0$ because it is zero at infinity. Thus, a flat front cannot describe realistically the front of the streamer which has a significant nonzero current given by equation (\ref{eq:J}). (This case is dealt with in Appendix~\ref{app:ns}.)

The system (\ref{eq:hydro_front}) for total current $J=0$ and with $\partial_t = -V\partial_\xi$ consists of the equations for $E$ and electron density $n$ in a flat front advancing with velocity $V$:
\begin{equation}
  \left.\begin{array}{rcl}
  \varepsilon_0 V \partial_\xi E &=& e v(E) n \\
  \partial_\xi \left\{[V\pm v(E)] n\right\} &=& -\nu_t(E) n
  \end{array}\right\}
  \label{eq:front}
\end{equation}
The photoionization term was neglected inside the front; it will be tackled in Subsection~\ref{ssec:photo}. 

The solution for $n(E)$ is obtained by dividing the second equation by the first and integrating:
\[ \frac{en(E)}{\varepsilon_0}=\frac{V}{V\pm v(E)}[\Psi(E_f)-\Psi(E)] \]
where we used $n=0$ at $E=E_f$ (corresponding to $\xi\rightarrow\infty$), and introduced the \emph{ionization integral} \citep[p.~62]{book:Lagarkov+Rutkevich1994}
\begin{equation}
  \Psi(E) = \int_0^E \alpha_t(E')\,d E'
  \label{eq:Psi}
\end{equation}
and $\alpha_t$ as the net spatial ionization rate (the inverse avalanche length, also called the first Townsend coefficient):
\begin{equation}
  \alpha_t(E)\equiv\frac{\nu_t(E)}{v(E)}=\frac{\nu_t(E)}{\mu(E) E}
  \label{eq:alpha}
\end{equation}
The field behind the front is necessarily $E(\xi=-\infty)=0$ for $J\equiv 0$, thus this system cannot produce $E_s\ne 0$. The ionization inside the flat front is therefore $n_s\equiv n(E=0) = \Psi(E_f)$. To be more accurate, however, we take $n_s\equiv n(E=E_s)$:
\begin{eqnarray}
  \frac{e n_s}{\varepsilon_0} &=& \frac{V}{V\pm v(E_s)}[\Psi(E_f)-\Psi(E_s)] \nonumber \\
     &\approx& \Psi(E_f)-\Psi(E_s)
  \label{eq:ns}
\end{eqnarray}
where we neglect $v(E_s)$ compared to $V$. The correction of $\Psi(E_s)$ is only necessary if $\Psi(E_s)$ is very different from zero. This may occur in a numerical implementation since the three-body attachment coefficient contributing to $\alpha_t$ makes it very large (and negative) for small $E$.

\subsubsection{Bulk charge relaxation time}
\label{sssec:relaxation}
One can make a ballpark estimation of $n_s$, e.g, for a power function $\alpha_t(E)\propto E^{k-1}$ used by some authors \citep[e.g.][ch.~3]{book:Bazelyan+Raizer1998}, with not too large $k$ (the mentioned reference used $k=2.5$):
\[ n_s = \frac{\varepsilon_0\nu_t(E_f)}{k e\mu(E_f)} \]
We can now support the claim from Subsection~\ref{ssec:general} that the Maxwellian relaxation time inside the streamer is small enough to be neglected. It is
\[ \tau_M = \frac{\varepsilon_0}{\sigma} = \frac{\varepsilon_0}{e\mu(E_s)n_s} = k\frac{\mu(E_f)}{\mu(E_s)}\tau_i\sim \tau_i \]
where $\tau_i = 1/\nu_t(E_f)$ is the ionization time in the streamer front. During this time, the streamer travels distance $V/\nu_t(E_f)$, which, as we will see later in equation (\ref{eq:d}), is the streamer front thickness $d\ll a$. Thus, the charges completely relax before streamer covers any significant distance.

The power-law dependence of $\alpha_t(E)$ was used only for an estimate, and any other sufficiently fast growing function would give a result of the same order of magnitude. Equality of Maxwellian relaxation time behind the front and the ionization time inside the front has been considered to be the condition for stable streamer propagation \citep{Pasko+1998}. We emphasize that $\tau_M$ is the time of relaxation of the bulk charge only; the surface charge relaxes at a much longer timescale \citep{Dyakonov+Kachorovskii1989}, which allows the existence of nonzero field $E_s$.

\subsubsection{Front thickness $d$}
When the front is completely flat, $E=E_f$ for all $\xi\gg d$. In this region, the second equation of (\ref{eq:front}) is:
\[ [V\pm \mu(E_f)E_f] \partial_\xi n = -\nu_t(E_f) n \]
This has an exponential solution $n\propto e^{-\xi/d}$ with
\begin{equation}
  d = \frac{V\pm v(E_f)}{\nu_t(E_f)}
  \label{eq:d}
\end{equation}

\subsubsection{Non-flat front (with $J\ne 0$)}
Now, let us add the effects of the finite transverse size of the streamer, in particular the current flowing along the streamer axis through the front, and estimate corrections to the expressions for $n_s$ and $E(\xi)$. The details of this procedure are too technical and therefore presented not here but in Appendix~\ref{app:ns}. Let us just recap the main results. In particular, we argue that we still can use expression (\ref{eq:ns}) for $n_s$ in which we still must use $E_f$ instead of $E_m$. We also obtain expressions for $\xi_m$ and $E_m$ (equation \ref{eq:app:Em}) which we repeat here for convenience:
\begin{equation}
  \frac{E_m}{E_f}=1-\frac{d}{l}\log\left(\frac{l}{d}\right),\quad \xi_m = d\log\left(\frac{l}{d}\right)
  \label{eq:Em}
\end{equation}
where the front thickness $d$ is given by equation (\ref{eq:d}). For a non-flat front, we may also use the corrected value $E_m$ instead of $E_f$ in expression for $d$.

\subsection{Relation between streamer velocity $V$ and its radius $a$ determined by photoionization}
\label{ssec:photo}

\subsubsection{\citet{Zheleznyak+1982} expression for photoionization source $s_p$}
Photoionization is proportional to the impact ionization $\nu_i n$:
\begin{equation}
   s_p(\rvec) = \int CF(\abs{\rvec-\rvec'})\nu_i(\abs{\Evec(\rvec')})n(\rvec')\,d^3\rvec'
   \label{eq:photo}
\end{equation}
where $C \approx 0.1\frac{p_q}{p+p_q}\sim 0.01$ (with $p=760$~mmHg, $p_q=30$--60~mmHg) \citep[as used in][]{Lehtinen+Ostgaard2018} and $F$ is the \citet{Zheleznyak+1982} function, which describes nonlocality due to photon transport:
\begin{equation}
  F(r) = \frac{e^{-r/\Lambda_2}-e^{-r/\Lambda_1}}{4\pi r^3\log(\Lambda_2/\Lambda_1)}
  \label{eq:zheleznyak}
\end{equation}
In the sea-level air $\Lambda_2 \approx 2$~mm and $\Lambda_1\approx 35$~$\mu$m and therefore $\log(\Lambda_2/\Lambda_1)\approx 4$.  The constant coefficient is such that $F$ is normalized to 1 when integrated over the whole space.

\subsubsection{Approximate expression for $s_p$}
Since most of the impact ionization occurs in a small region $\abs{\xi}\lesssim d$, and the photoionization electrons are created at $\xi\sim a\gg d$ (see Subsection~\ref{ssec:photodistance}), for calculation of $s_p$ we may assume that the source of photons is at $\xi=0$. The impact ionization integrated over the whole volume is
\[ S_i = \pi\aph^2 \int \nu_i(E) n \,d\xi \]
Here $\aph$ is the effective radius of the photon-emitting portion of the front, or \emph{photoemitting radius} for short. In most of the results in this paper we take $\aph=a/2$. The integral
\[ \int \nu_i(E) n \,d\xi \approx n_s [V \pm v(E_s)] \]
may be obtained by integrating the second equation of (\ref{eq:front}) and taking $\nu_i(E) \approx \nu_t(E)$. Usually, $v(E_s)\ll V$, but for now we will keep this term.

Substituting this into the photoionization source (\ref{eq:photo}), we get:
\begin{equation}
  s_p(\xi) = S_i C\Feff(\xi) = \pi\aph^2 n_s [V \pm v(E_s)] C \Feff(\xi)
  \label{eq:s_p}
\end{equation}
where function $\Feff(\xi)$ is $F(r)$ averaged over the transverse area of the front. This averaging is necessary to avoid divergence in $F$ as $\xi\rightarrow 0$ and it takes into account that at $\xi\lesssim\aph$, the transverse size of the emitting region $\sim\aph$ also plays a role:
\begin{eqnarray*}
  \Feff(\xi) &=& \frac{1}{\pi\aph^2}\int_{r_\perp<\aph}F(r)\,d^2\rvec_\perp \\
  &=& \frac{1}{\pi\aph^2}\int_\xi^{\sqrt{\xi^2+\aph^2}}F(r)\,2\pi r\,d r,\quad r = \sqrt{\xi^2+r_\perp^2}
\end{eqnarray*}
This may be calculated numerically, using (\ref{eq:zheleznyak}).

\subsubsection{\citet{Pancheshnyi+2001} equation}
The second equation of (\ref{eq:front}) with added photoionization source (or the second equation of (\ref{eq:hydro_front}) taken on the streamer axis, see also equation (\ref{eq:app:front_with_current}) in the Appendix~\ref{app:ns} where we consider a non-flat front) is
\begin{equation}
  -\partial_\xi\left[(V\pm v)n\right] = \nu_t n + s_p(\xi)
  \label{eq:n_eq_photo}
\end{equation}
It is a linear inhomogeneous first-order differential equation. With boundary condition $n(+\infty)=0$, its solution is
\begin{equation}
 n(\xi) = \frac{1}{V\pm v(E)}\int_\xi^\infty s_p(\xi')\exp{\left[\int_{\xi}^{\xi'}\frac{\nu_t(E)\,d\xi''}{V\pm v(E)}\right]}\,d\xi'
\label{eq:nphoto}
\end{equation}
Substituting (\ref{eq:s_p}), from the condition $n(0)=n_s$ we get an equation for electron density balance first derived by \citet[equation~(17)]{Pancheshnyi+2001}:
\begin{equation}
  \pi\aph^2C\int_0^\infty\Feff(\xi)\exp\left[\int_0^\xi\frac{\nu_t(E)\,d\xi'}{V \pm v(E)}\right]\,d\xi = 1
  \label{eq:photobalance}
\end{equation}

\subsubsection{Approximate forms of \citet{Pancheshnyi+2001} equation}
\citet{Naidis2009} used a boundary condition on $n(\xi)$ at a fixed distance $n(\xi_p)=n_p$ to describe the photoionization source instead of using a distributed source $s_p(\xi)$ and thereby obtained an approximate form of this equation. This may be done by setting $\Feff(\xi)=0$ for $\xi<\xi_p$ in (\ref{eq:n_eq_photo}) plus the mentioned boundary condition, which gives the solution
\[
  \log\left[\frac{(V\pm v[E_s])n_s}{(V\pm v[E(\xi_p)])n_p}\right] = \int_0^{\xi_p}\frac{\nu_t(E)\,d\xi'}{V \pm v(E)}
\]
The value of $n_p$ may be calculated from (\ref{eq:nphoto}) at $\xi=\xi_p$. An even more approximate form was obtained first by \citet{Loeb1965} by  neglecting the drift velocity and taking $\nu_t\approx\const$:
\begin{equation}
 V \approx \frac{1}{\log(n_s/n_p)}\int_0^{\xi_p}\nu_t(\xi)\,d\xi \approx \frac{\nu_t \xi_p}{\log(n_s/n_p)}
 \label{eq:Loeb65}
\end{equation}
Of course, with such an approximate approach a question remains of how to determine $\xi_p$ (and $n_p$). The results of Subsection~\ref{ssec:photodistance} indicate that $\xi_p\sim\aph$.

\section{Application to streamers in dry air at sea level}
\label{sec:dispersion}
\subsection{The expressions for physics coefficients}
\label{ssec:air_coefs}
To avoid possible discontinuities in the solution, we avoided using piecewise-approximated coefficients given, e.g., by \citet{Morrow+Lowke1997}. Instead, we use the following smooth but approximate expressions.
\begin{enumerate}
\item The electron mobility:
\[ \mu(E) = \left(\frac{E}{\sci{3}{6}\mbox{~V/m}}\right)^{-0.17}\times0.044\mbox{~m$^2$\,V$^{-1}$\,s$^{-1}$} \]
This expression was chosen because it fits well the piecewise expression of \citet{Morrow+Lowke1997}. Note that this gives an infinite conductivity as $E\rightarrow 0$. However, it may be estimated that this expression holds well for $E\gtrsim 30$~kV/m, in which range all the fields that we consider belong. The power law coefficient is close to value of $-0.2$ suggested by \citet{Babaeva+Naidis1997}.
\item The ionization rate is fitted to \citet{Morrow+Lowke1997} using the Townsend approximation:
\begin{equation}
   \nu_i(E) = v(E)\alpha_{i0}e^{-E_{i0}/E}
   \label{eq:nu_i}
\end{equation}
where
\[ \alpha_{i0}=\sci{5.4}{5}\mbox{~m$^{-1}$},\ E_{i0}=\sci{1.95}{7}\mbox{~V/m} \]
This gives smaller values at lower fields (reduction at about 30\% at $E=3$~MV/m) but we have a reason to believe that the streamer propagation is determined mostly by conditions at the head, characterized by much higher fields $\sim E_m$. Moreover, at even lower fields we expect the attachment to dominate.
\item The attachment rate $\nu_a(E)=v(E)(\alpha_{a2}+\alpha_{a3})$ is the sum of the two-body and three-body rates. The two-body spatial attachment rate is fitted to \citet{book:Pasko1996} at $E<3.5$~MV/m using the Townsend approximation:
\[ \alpha_{a2}(E)=\alpha_{a0}e^{-E_{a0}/E} \]
\[ \alpha_{a0}=\sci{1.7}{3}\mbox{~m$^{-1}$},\quad E_{a0}=\sci{3.2}{6}\mbox{~V/m} \]
This also gives a $\sim$30\% lower value than both \citep{book:Pasko1996} and \citep{Morrow+Lowke1997} at fields $E=2$--2.5~MV/m, however, together with the reduced value of $\nu_i$ given by (\ref{eq:nu_i}) gives an approximately correct interception field $E_k \approx 2.86$~MV/m. This field is defined as $\nu_a(E_k)=\nu_i(E_k)$ or $\nu_t(E_k)=0$ ($\nu_t$ is defined in item \ref{item:net} below). The theoretical values for $E_k$ are: 2.8~MV/m from $\nu_{i,a}(E)$ in \citep{Morrow+Lowke1997}, 3.0~MV/m from our simulation with BOLSIG$+$ \citep{Hagelaar+Pitchford2005}, or 3.1~MV/m \citep[p.~26]{book:Bazelyan+Raizer1998}; the experimental value is 2.4--2.6~MV/m \citep[p.~338]{book:Raizer1991}, but this may be underestimated because the attachment may have been compensated by detachment.

The three-body spatial attachment rate is important at low fields $E\lesssim 1$~MV/m which are typical for the streamer channel interior, and therefore are necessary to consider for the results of Subsection~\ref{ssec:att}. It is taken from \citep{Morrow+Lowke1997}:
\begin{eqnarray*}
 \alpha_{a3}(E) &=& \sci{4.7778}{-69}\times N^2 \times \\
  & &\left[E/N\times 10^4\right]^{-1.2749} \mbox{~m$^{-1}$}
\end{eqnarray*}
where $E$ is in V/m and $N=\sci{2.688}{25}$~m$^{-3}$ is the neutral density (Loschmidt constant).

The effects of attachment at the streamer front, however, were shown not to be very important. We performed exactly the same calculations without attachment, and the calculated streamer parameters turned out not to be very close to when it was included.
\item \label{item:net} The temporal and spatial net ionization rates are defined as
\[ \begin{array}{rcl} \nu_t(E) &=& \nu_i(E)-\nu_a(E) \\ \alpha_t(E)&=&\alpha_i(E)-\alpha_a(E) \end{array} \]
\end{enumerate}

\subsection{Procedure of finding $V$ and other streamer parameters}
\label{ssec:procedure}
Now, let us collect everything together. Let us fix only the external field $E_e$ (determined by the experimental setup) and streamer length $L$ (determined by the previous propagation history). The unknown streamer parameters are $a$, $V$, $n_s$, $E_s$, $E_m$. Parameter $E_m$ is expressed by equation (\ref{eq:Em}) in terms of $E_f$, which is more convenient to use as a parameter in the solution process. We have the following relations between parameters:
\begin{enumerate}
\item $E_f$ depends on the channel geometry and $E_s$ by equations (\ref{eq:E}) for $E(\xi)$ and (\ref{eq:Ef}). It describes the external field enhancement by the conducting streamer channel.
\item The field inside the streamer $E_s$ is related to $V$ and $n_s$ by the current continuity equation (\ref{eq:Jcontinuity}).
\item Streamer electron density $n_s$ is fixed by the flat-front theory equation (\ref{eq:ns}) and is a function of $E_f$ (and also weakly depends on $E_s$). It may be interpreted as the condition for stable streamer propagation, as discussed in Subsubsection~\ref{sssec:relaxation}.
\item Velocity $V$ is related to radius $a$ by equation (\ref{eq:photobalance}), which fixes electron density balance in the photoionization and impact ionization processes.
\end{enumerate}

Thus, the problem of the streamer description had been reduced to a system of algebraic equations (SAE). We repeat these equations here for convenience:
\begin{equation}
\def\arraystretch{2.2}
\left.\begin{array}{c}
  \displaystyle E_f = E_e + (E_e-E_s) \eta \\
  \displaystyle e n_s v(E_s) = \frac{\varepsilon_0 V (E_e-E_s) \eta}{l} \\
  \displaystyle \frac{e n_s}{\varepsilon_0} = \int_{E_s}^{E_f}\alpha_t(E)\,d E \\
  \displaystyle \pi\aph^2C\int_0^\infty\Feff(\xi)\exp\left[\int_0^\xi\frac{\nu_t(E)\,d\xi'}{V \pm v(E)}\right]\,d\xi = 1
\end{array}\right\}
\label{eq:system}
\end{equation}
Here $\eta,l$ are functions of $L/a$ given by equations (\ref{eq:eta}) and (\ref{eq:l}), respectively; $v(E)=\mu(E)E$ is the electron drift velocity; $\nu_t(E)$, $\alpha_t(E)$ and $\mu(E)$ are defined in Subsection~\ref{ssec:air_coefs}; function $\Feff(\xi)$, constant $C$ and the photoemitting radius $\aph$ (which may be taken as a fixed fraction of $a$, e.g., $\aph=a/2$) are defined in Subsection~\ref{ssec:photo}. The laboratory parameters (given) are $E_e$ and $L$, the unknown parameters are $a$, $V$, $E_s$, $n_s$ and either $E_f$ or $E_m$ expressed by equation (\ref{eq:Em}). Finally, the field as a function of coordinate in the last equation is given by equation (\ref{eq:E}):
\begin{equation}
  E(\xi'>0) = E_e + \frac{E_f-E_e}{1+\xi'/l}
  \label{eq:system_E}
\end{equation}
where we took $S\equiv\Omega$, the step function as the switch function. It is possible to use expression (\ref{eq:app:S}) derived in Appendix~\ref{app:ns}, which is more computationally intensive but the solution does not change significantly.

There are four equations but five unknowns. We can quickly reduce the number of equations if we substitute $E_f$ from the first and $n_s$ from the third equation to into the other two equations, in which case we are left with two equations and three unknowns $a$, $V$, $E_s$. Still, there is one unconstrained unknown. It is most convenient to choose the streamer radius $a$ to be the independent variable. This means that we still cannot fix all the streamer parameters based only on $E_e$ and $L$: the radius $a$ is not fixed. From the SAE (\ref{eq:system}) we may find $V$ and other parameters as functions of $a$. We may call each solution at given $a$ a ``streamer mode'' and the functional relations $V=V(a)$, etc., the streamer ``dispersion equation'' by analogy with the modes of flat-front perturbations with different transverse wavenumbers \citep{Derks+2008}. This analogy was mentioned in Introduction and is further discussed below in Subsection~\ref{ssec:flatfront}.

The computational algorithm to solve SAE (\ref{eq:system}) exploits the fact that the intrinsic field $E_s$ lies in a finite range, namely it never exceeds $E_e$. We proceed as following for each given $a$:
\begin{enumerate}
\item For all possible $E_s$ in the valid range $[0,E_e]$, we calculate $E_f$ and $n_s$ using the first and the third equation, respectively.
\item The result for $E_f$ is used in expression (\ref{eq:system_E}) for $E(\xi')$ which is substituted into the left-hand side of the fourth equation. We use the fact that it is a monotonously decreasing function of $V$ and use binary search to find $V$ at which it is equal to 1.
\item This value of $V$, together with $n_s$ from step 1, is used in the second equation to calculate a new value of the intrinsic streamer field
\[ E^\mathrm{new}_s = \frac{\varepsilon_0 V\eta (E_e-E_s)}{e n_s \mu(E_s)l} \]
\item We locate the set of values of $E_s$ inside the range $[0, E_e]$ such that $E^\mathrm{new}_s=E_s$.
\item The found value(s) of $E_s$ are used to calculate all the other parameters, e.g. $V$ in step 2. This is the solution of the system. Each unique set of parameters $(E_s,V,\dots)$ gives the ``streamer mode'' structure corresponding to the given value $a$.
\end{enumerate}
\begin{figure}
\includegraphics[width=0.49\textwidth]{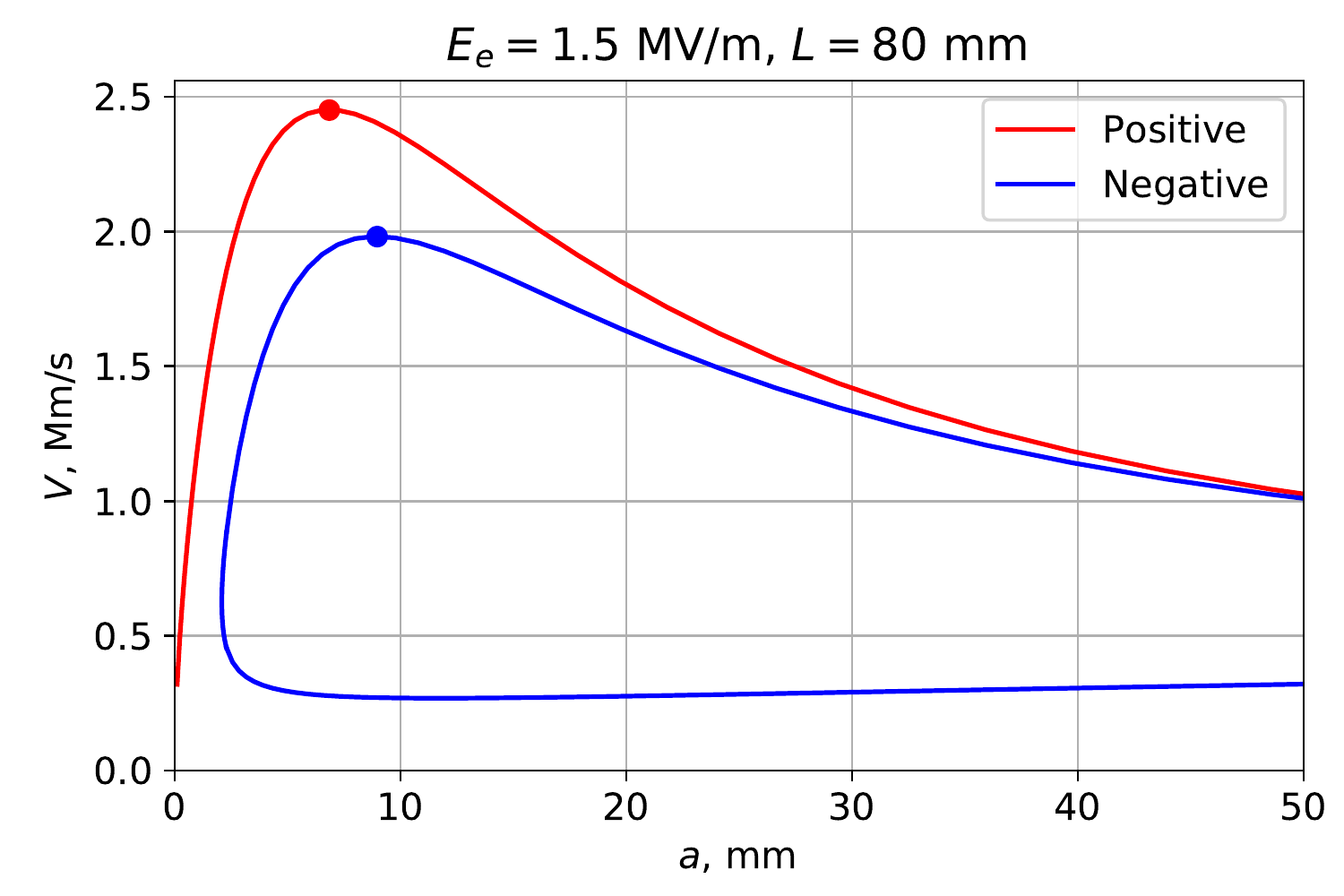}
\caption{An example of a set of streamer solutions $V(a)$ (the streamer dispersion function). The large dot indicates the values $a^\ast$ and $V^\ast\equiv\max V$ at which $V$ is maximized.}
\label{fig:branches}
\end{figure}
An example of the result of an application of such algorithm is shown in Figure~\ref{fig:branches}. We see that the dispersion function $V(a)$ calculated in this way is not always unique, but sometimes has multiple branches, like the shown solution for negative streamers. The unique solutions for streamer parameters will be obtained on the basis of max-$V$ criterion introduced in Section~\ref{sec:maxv} and presented Section~\ref{sec:all_results}.

\subsection{Results for parameters as functions of $a$}
We performed calculations for both positive and negative streamers with lengths in interval $L=5\dots 200$~mm with step 5~mm and external fields in interval $E_e=0.1\dots 3$~MV/m with step 0.05~MV/m. (Not all results are presented in this paper, but are available on request, or may be reproduced by the reader with the help of attached software, see Appendix~\ref{app:code}.) In this Section, we only present the dispersion functions for positive streamers, $L=120$~mm and $E_e=0.4$, 0.6, 0.8, and 1~MV/m.

\begin{figure}
\includegraphics[width=0.49\textwidth]{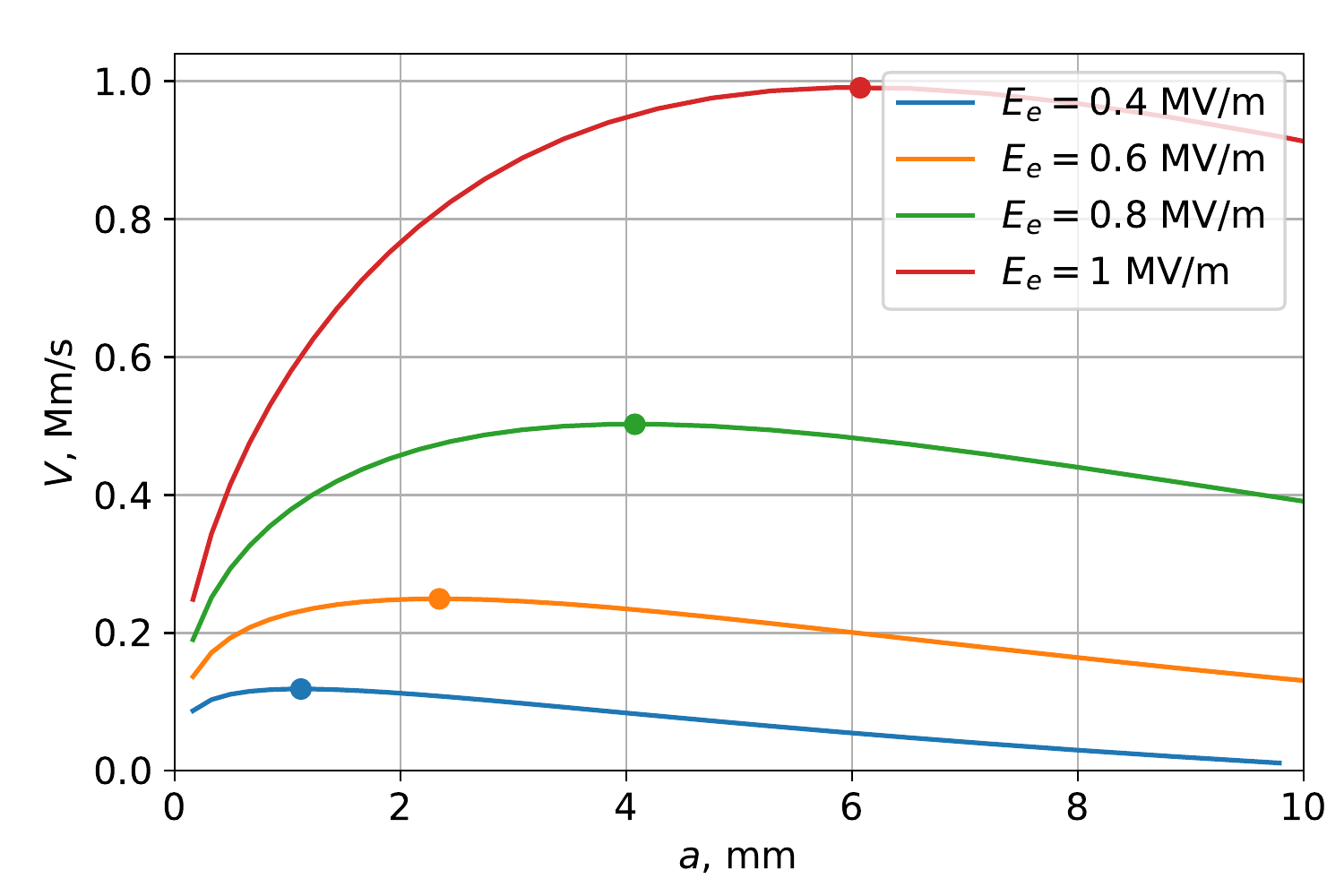}
\caption{Streamer velocity $V$ as a function of streamer radius $a$ (dispersion function), for positive streamers of length $L=120$~mm and a set of selected values of external field $E_e$. The large dot indicates the values $a^\ast$ and $V^\ast\equiv\max V$ at which $V$ is maximized.}
\label{fig:a_results_v}
\end{figure}

\begin{figure*}
\includegraphics[width=0.98\textwidth]{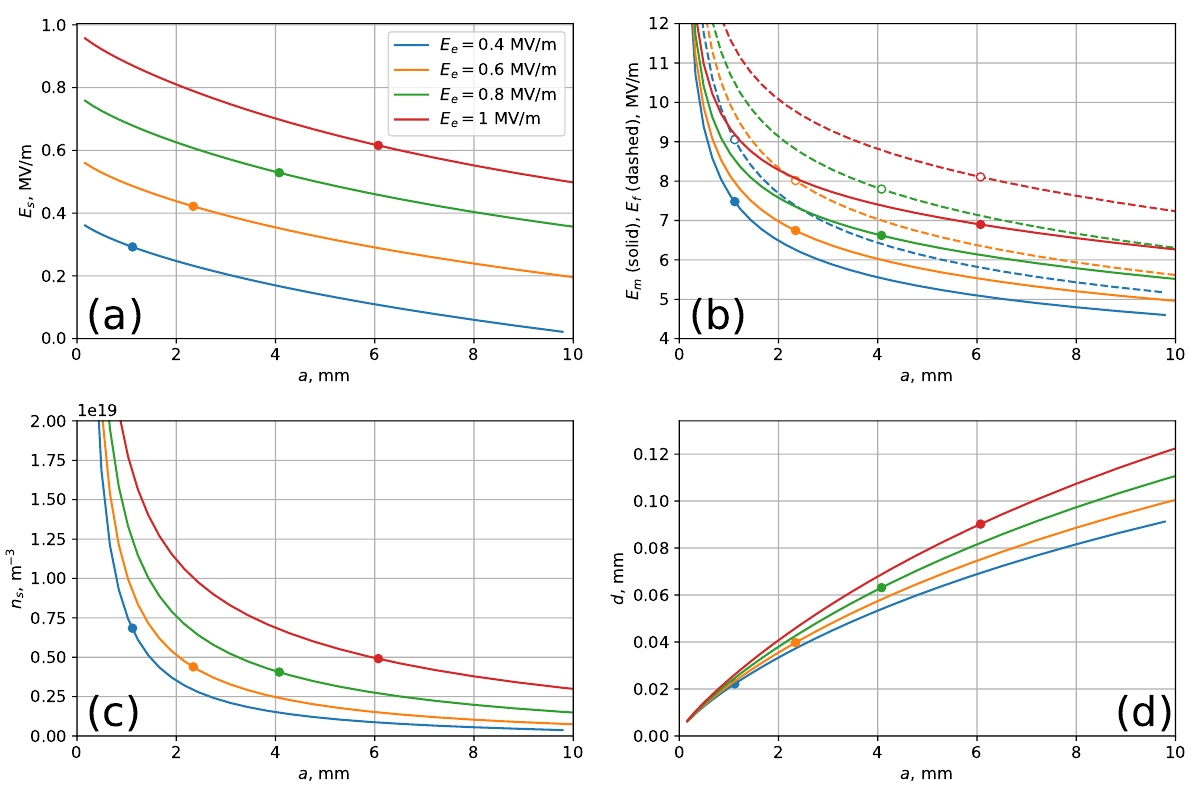} 
\caption{(a)~Streamer intrinsic field $E_s$, (b)~ionization $n_s$, (c)~maximum field $E_m$ ($E_f$), and (d)~ionization front thickness $d$ as functions of streamer radius $a$, for positive streamers of length $L=120$~mm and a set of selected values of external field $E_e$. The large dot indicates the value $a^\ast$ at which $V$ is maximized.}
\label{fig:a_results}
\end{figure*}

The calculated values of streamer velocity $V$ are presented in Figure~\ref{fig:a_results_v}, and the intrinsic field $E_s$, ionization $n_s$, maximum field $E_m$ ($E_f$) and ionization front thickness $d$ are presented in Figure~\ref{fig:a_results}. For the streamer velocity $V$, it is convenient to use a unit of Mm/s $=$ mm/ns. We immediately observe that most of the streamer parameters are monotonic functions of $a$, except the streamer velocity $V$ which has a maximum at a certain value of $a=a^\ast$. This value is highlighted with a large dot in each Figure.

In particular, in Figure~\ref{fig:a_results}, we can observe that there is a significant difference between the idealized maximum field $E_f$ and the actual one $E_m$. This is due to the fact that in equation (\ref{eq:Em}), even though $d/l\ll 1$, the value of $\log(l/d)\gg 1$ so that the ratio $E_m/E_f$ can be as low as 0.8.

\subsection{Distance of photoionization}
\label{ssec:photodistance}
\begin{figure*}
\includegraphics[width=0.98\textwidth]{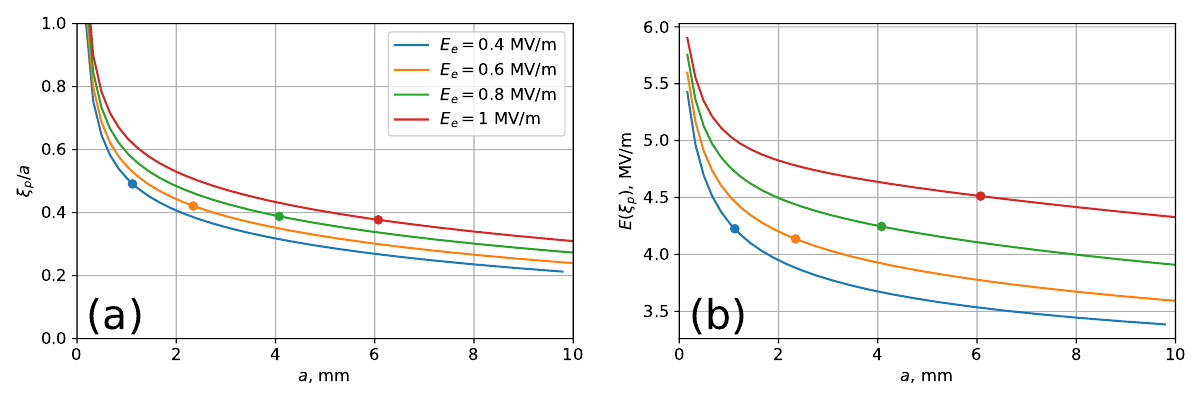} 
\caption{(a)~The photoelectron seed distance $\xi_p$ and (b)~the field at that location $E(\xi_p)$, for positive streamers of length $L = 120$~mm and a set of selected values of external field $E_e$, with $\aph=a/2$. The large dot highlights the value of $a=a^\ast$ at which $V$ is maximized.}
\label{fig:maxphoto}
\end{figure*}

Many researchers \citep{Gallimberti1972,Naidis2009}, starting with the classical paper of \citet{Dawson+Winn1965}, assumed that the seed photoelectrons are produced at the distance $\xi_p$ which is determined by the condition $E=E_k$, where $E_k$ is the critical field at which $\nu_t(E_k)=0$ ($E_k\sim 2.86$~MV/m for the values used here). However, \citet{Babaeva+Naidis1997} suggested that it is of the order of streamer radius, $\xi_p\sim a$. We can give an accurate answer to this question by finding at which $\xi$ the expression under the integral in equation (\ref{eq:photobalance}) gives the biggest contribution to the integral. For example, we may take the median of that expression. The values for $\xi_p$ and $E(\xi_p)$ that were found this way are plotted in Figure~\ref{fig:maxphoto}. We see that, in fact, the photoelectron seed distance is determined by streamer radius, or more precisely, is approximately equal to the photoemitting radius which was assumed to be $\aph=a/2$ in the presented calculations. The corresponding electric field is also significantly higher than $E_k$.

\section{Necessity of max-$V$ criterion}
\label{sec:maxv}

We have solved the SAE (\ref{eq:system}) to find the streamer modes, namely $V$, $E_s$, $n_s$ and $E_f$ as functions of $a$. In the results presented in Figures~\ref{fig:a_results_v}--\ref{fig:a_results}, we notice that the streamer velocity $V$ maximizes at a chosen value $a=a^\ast$, while all other parameters are monotonous in $a$. Incidentally, the fact that $V(a)$ has a maximum is equivalent to a statement that $a(V)$ has two solutions for $V<V_\mathrm{max}=V(a^\ast)\equiv V^\ast$, which was noticed by \citet{Pancheshnyi+2001} and demonstrated in their Figure~9.

The value of $a^\ast$ seems to be special and we may hypothesize that it is also the correct one at which the streamer actually propagates, i.e., the ``choice'' is made by maximizing velocity $V$. This criterion (which we will call max-$V$ criterion in this paper) may be used as the missing constraint which now allows for finding an unambiguous set of streamer parameters. We emphasize that max-$V$ is a \emph{heuristic} criterion because we do not have a strict mathematical basis for it at the present time; it only makes an intuitive sense that the ``true'' streamer ``wins a competition'' when we measure which ionization protrusions have reached a certain distance, because the fastest one gets there first. However, we will now advance some physics-based arguments in its favor, by developing analogy with flat-front perturbations.

\subsection{Analogy with the flat-front perturbation analysis}
\label{ssec:flatfront}
Let us draw an analogy between a streamer and a linear instability of a flat front \citep{Arrayas+Ebert2004,Derks+2008}. In a flat-front perturbation theory, the small changes in all physical values satisfy a system of linear partial differential equations. As a consequence, the transversely-harmonic perturbations of the flat front exponentially decrease or grow in time. (In other physical systems described by flat-front perturbations, it may be possible to have harmonically-oscillating solutions, e.g., waves on the water surface.) The state of the system at arbitrary moment $t>0$ is determined not only by the dynamic equations, which determine the time evolution, but also by the initial conditions at $t=0$. Let us start with a harmonic perturbation at transverse wavenumber $k$ and the initial amplitude of the front position being $\zeta^0_k$. Then the time evolution of the relative (in respect to the average, or unperturbed) front position $\zeta$ is given by
\[ \zeta(r_\perp,t) = \Real\left\{ \zeta^0_k e^{i k r_\perp + \gamma(k) t}\right\} \]
where $\gamma(k)$ is a growth (if $>0$) or decay (if $<0$) rate. (Depending on the nature of the front, multiple modes may be possible for the same $k$, but for simplicity let us choose only one of them, e.g., the fastest-growing.) The foremost part of a front is ahead of the average front position by
\[ L \equiv  \max\zeta(r_\perp,t) = \abs{\zeta^0_k} \exp[\gamma(k) t] \]
where we chose to denote it with $L$ in order to keep in mind that it is analogous to the length of a streamer in our problem. The velocity of the front is:
\[ v(r_\perp,t) = V_0+\frac{d\zeta}{d t} = V_0+\Real\left\{ \gamma(k) \zeta^0_k \exp[i k r_\perp + \gamma(k) t]\right\} \]
where $V_0$ is the velocity of the unperturbed (flat) front. At time $t$ when protrusion reaches $L$, the protrusion velocity (i.e., at the same $r_\perp$ as $\zeta=L$) is
\[ V(L,k) = V_0+ \gamma(k) \abs{\zeta^0_k} \exp[\gamma(k) t] = V_0 + \gamma(k) L \]
If the initial conditions are random fluctuations which is a linear combination of harmonic perturbations at all possible $k$, then at some advanced time, independently of the initial values $\zeta^0_k$, only the protrusion with the highest growth rate
\[ \gamma^\ast = \max_k \gamma(k) = \gamma(k^\ast) \]
will survive. If we choose a fairly large $L$ ($\gg\zeta^0_k$), so that the above-defined ``advanced time'' is elapsed when it is reached, chances are that it is done by the mode $k^\ast$. Thus the ``true'' protrusion velocity, which is ``independent'' of the initial conditions, is
\begin{equation}
  V^\ast(L) = V(L,k^\ast) =  V_0 + \gamma^\ast L = \max_k V(L,k)
  \label{eq:maxVflat}
\end{equation}
We took ``independent'' into quotes because, strictly speaking, all other modes are still present, but their amplitudes do not contribute much. Maximizing over $k$ is equivalent, of course, to maximizing over the transverse size $a=1/k$, which is analogous to the streamer radius. By analogy with the flat-front perturbation theory, we may call the dependencies $V(a)$ which we have obtained in Section~\ref{sec:dispersion} the streamer ``dispersion functions'' and the sets of solutions for each $a$ the streamer ``mode structure.''

Thus, we have shown that the preferred solution for the protrusion of a flat front which is grown to amplitude $L$ also has the highest velocity out of all solutions with transverse sizes $a$ and same $L$. One may argue that if this argument is valid for small perturbations only, it is not necessarily valid for a streamer, which is a large perturbation. However, this invalidity must start only at a certain value of $L$ when the perturbation becomes sufficiently nonlinear, so this argument may continue to be approximately valid.

Hydrodynamic simulations are a popular method of numerical solution of microscopic equations \citep{Dhali+Williams1987,Morrow+Lowke1997,Teunissen+Ebert2017}. When one simulates a flat front, the preferred solution automatically arises because of small numerical fluctuations \citep{Derks+2008}. Therefore, we may state that in the case of nonlinear streamer simulations of a streamer discharge, the preferred solution with the fixed radius $a^\ast$ will also arise automatically because of fluctuations in initial conditions and numerical fluctuations in the course of calculations.

\subsection{Necessity of max-$V$ or a similar criterion}
One might ask whether the system (\ref{eq:system}) is really the maximum number of equations available. If we missed one due to lack of knowledge, the missing equation could have fixed the remaining free parameter $a$.  However, as we just demonstrated, in a flat-front theory the behavior is determined by initial conditions. If we start with a single mode without any other fluctuations, then $k$ (or equivalently, $a=1/k$) is given by initial conditions only, and thus is a free parameter. Therefore, not only the transverse size $a=1/k$ in the case of the flat-front perturbations, but by extension also \emph{the radius of the streamer cannot be obtained on the basis of microscopic equations such as (\ref{eq:hydro}) alone in principle}. All solutions with different $a$ are valid solutions. The only way a certain transverse size $a^\ast$ can emerge is if there are multiple random small fluctuations present in the beginning, then only one of them survives in the long run. As in the flat-front perturbations theory, the correct solution is not a stable solution, but the most unstable one.

\subsection{Invalidity of extremization of an extensive physical value as an alternative to max-$V$ criterion}
In classical mechanics, motion of a conservative system is determined by the principle of least action. Action is an extensive physical value, i.e. is obtained by integrating over volume of the system and thus is (approximately) proportional to the system size. We obviously cannot apply the minimization of action to a streamer system because application of this principle still requires the initial conditions and we just demonstrated that in the absence of fluctuations the streamer radius is determined by the initial conditions and does not have an unambiguous value. However, one might argue that it may be possible to find an extensive variable which would be maximized for the ``correct'' streamer radius. For example, the faster the streamer grows, the faster the electric energy is converted into kinetic energy of free electrons so we may expect some correlation (if not direct correspondence) between this energy conversion rate and the velocity of the streamer and maybe look to replace the max-$V$ criterion with maximization of this rate. Energy balance during the streamer propagation was considered, e.g., by \citet{Gallimberti1972}, although he did not look into its maximization. We calculated this rate in our model, but it did not exhibit a maximum in $a$. This is due to the fact that conversion rate is also proportional to the area of the streamer head, so it grows indefinitely with the radius.

In fact, we may advance a general argument against extremization of any extensive physical value as a criterion.

Consider a translationally-symmetric system such as one analyzed in this paper, namely infinite flat parallel electrodes (or just one electrode if the other is sufficiently far away, as we considered) that create a uniform electric field in the space between them. In this system, we can have a single streamer (as we considered), but we can also have multiple identical streamers separated transversely by a distance large enough that they do not affect each others' propagation. A system with any number of streamers satisfies the same microscopic equations such as (\ref{eq:hydro}). An extensive parameter, such as the total energy conversion rate, will be proportional to the number of streamers, and may be increased by simply adding more streamers while keeping the parameter values constant. Thus, a system in which it attains a maximum value cannot possibly describe a single streamer.

Another example of an extensive physical value that is invalidated by this argument is the total current flowing between electrodes while the streamer is propagating.

\subsection{Viable alternatives to max-$V$ criterion}
We still do not exclude a possibility that maximization of an \emph{intensive} physical value may be an alternative to the max-$V$ criterion. However, this alternative criterion must be equivalent to max-$V$ in the case of small perturbations of a flat front. As an example (although for a completely different physical system), we may quote the max-amplitude criterion of \citet{Dias+Miranda2013} who considered ``fingering'' (Saffman-Taylor) instability in viscous flows. They still used linear approximation to calculate the perturbation growth, but the growth rate (analog of $\gamma$ above) was time-dependent, which necessitated the new criterion in their problem. In application to our system, this criterion is equivalent to minimizing the time required for the streamer to reach a given length $L$. But this can be achieved only by maximizing velocity at each previous moment of time, i.e. at lengths $<L$, so the max-$V$ criterion must emerge again.

\subsection{Other authors' approach to the insufficiency of microscopic equations for determining parameters}
\label{ssec:other_authors}
\citet[p.~277]{book:Bazelyan+Raizer1998} approached the problem of missing one equation by arbitrarily fixing one streamer parameter, namely the electric field in front of the streamer $E_m$, at a ``chosen'' value. It was done on the basis of a consideration that $\nu_i(E)$ given by equation (\ref{eq:nu_i}) has a threshold-like behavior $\propto(E-E^\ast)$ where $E^\ast$ is the field at the ``neck'' of the streamer. Then $E_m\propto E^\ast$, and is expressed as a fixed fraction of $E_{i0}\approx 19.5$~MV/m used in the expression for $\nu_i$. Let us look at variations of $E_m$ plotted in Figure~\ref{fig:a_results}. We see that even though $E_m$ varies in a wide range for various $a$, but when taken at $a^\ast$ (highlighted with a dot), it is rather constant. This results is even more apparent in Figure~\ref{fig:pos_results}d, where the full set of results is presented only for the maximized $V$. However, unlike the ``chosen $E_m$'' hypothesis, there is a reason for it attaining this particular approximately constant value: for the set of parameters to which it belongs, the velocity is maximized. Moreover, we will see in Subsection~\ref{ssec:neg_results} and Figure~\ref{fig:neg_results}d that for negative streamers, ``chosen $E_m$'' is not constant at all. Moreover, the ``chosen $E_m$'' theory does not allow calculation of streamer parameters if the system does not have a threshold-like $\nu_i$. The proposed max-$V$ criterion, however, is universal in the sense that it does not depend on the exact functional shapes of the coefficients, such as $\nu_i$.

Even though fixing $E_m$ already determines all parameters, \citet{Dyakonov+Kachorovskii1989}, on whose model \citet[p.~277]{book:Bazelyan+Raizer1998} based theirs, actually overdetermined their system of equations by fixing also $E_s=E_e$ in their Section 4. This led to nonsensical results such that the ionized region in an infinitely growing streamer has finite length, even in the absence of attachment. Such logical errors are hard to locate and lead to obstruction of understanding and consequently general skepticism to the class of such ``simplified'' models in scientific community. We must point out that \citet{Dyakonov+Kachorovskii1989} also wrote out a complete system of equations corresponding to our equations (\ref{eq:E},\ref{eq:Jcontinuity},\ref{eq:ns},\ref{eq:photobalance}). In their paper, the corresponding equation numbers are (17, 19, 14, 9), even though these equations were obtained with different approximations. In their earlier paper, \citet[Figure~1]{Dyakonov+Kachorovskii1988} even discussed that changing the radius of the streamer to a more physical one may increase the velocity, but did not fully formulate the max-$V$ principle.

\subsection{Unambiguity of streamer parameters in our calculation}
The system (\ref{eq:system}) plus max-$V$ criterion unambiguously determined all parameters of the streamer from given $L$ and $E_e$. We emphasize again that $E_e$ is determined by the laboratory conditions and $L$ by the previous history of the streamer propagation, so these parameters are presumed to be known. After fixing $a=a^\ast$, we do not have any free parameters left in the model, except one dimensionless parameters $\aph/a=1/2$. We tried setting this parameter to 1, and $a^\ast$ only changed by a few percent, and the change in $V^\ast$ was even smaller. Thus, the method of determination of streamer parameters outlined in this paper is stable in respect variations of $\aph/a$, and hopefully may be applied for quantitative estimates in practical situations.

\section{Results with max-$V$ criterion applied}
\label{sec:all_results}

Now, we present results of calculations with the streamer radius selected at value $a^\ast$ by the max-$V$ criterion. In the presentation of these results, we omit the asterisk in the notation of the optimal value (i.e. write $a$ instead of $a^\ast$ etc.), hopefully this does not cause a confusion. In Appendix~\ref{app:analytic}, by making additional approximations, we derive analytic expressions for $V(a)$ and the optimal values $a^\ast$ and $V^\ast$ and explain the qualitative behavior of streamer parameters. However, the analytic expressions are less accurate and therefore cannot be used for quantitative estimates.

\subsection{Positive streamers}
\label{ssec:pos_results}
\begin{figure*}
\includegraphics[width=0.98\textwidth]{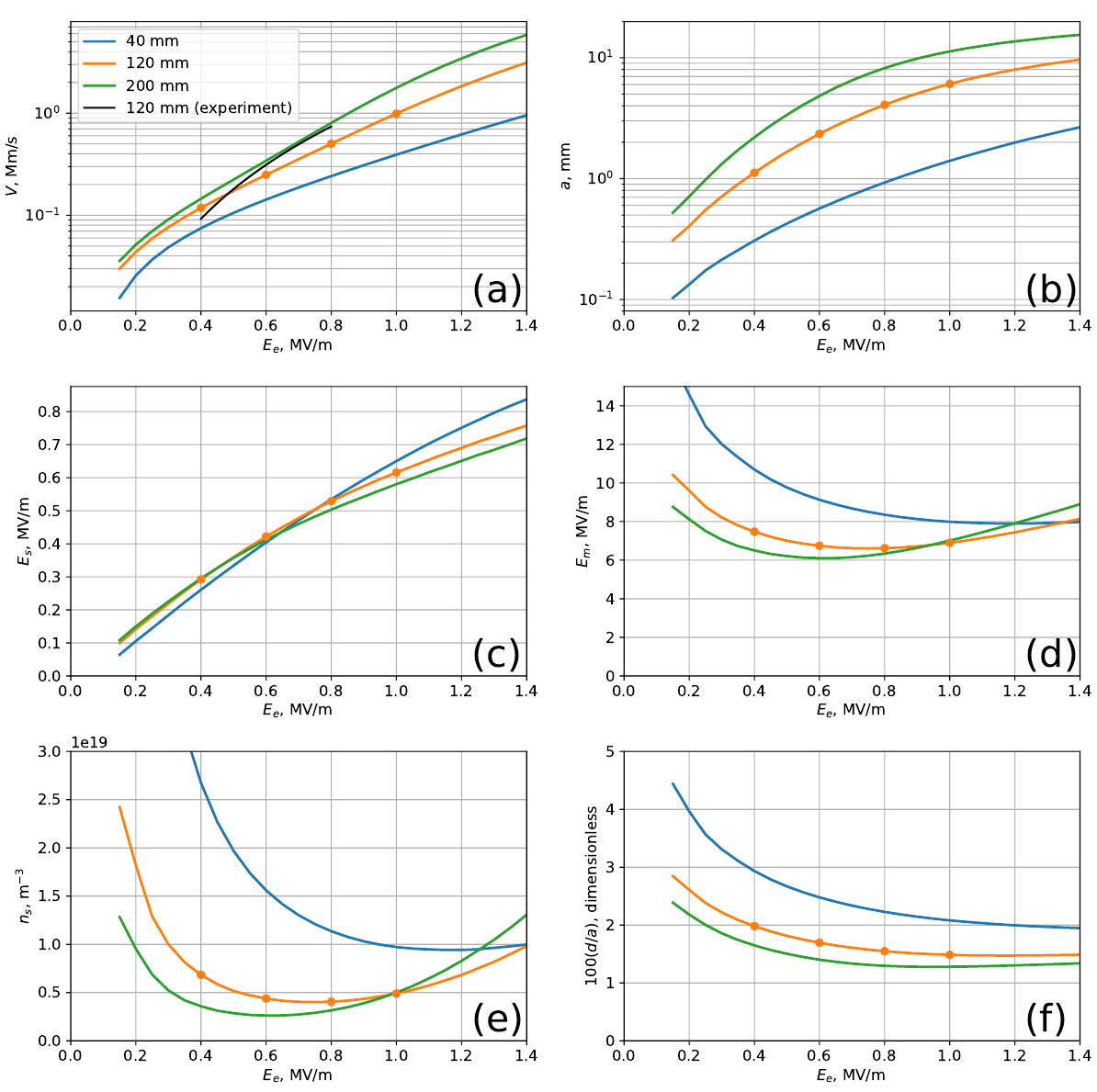}
\caption{Results for positive streamers (at maximum $V$) as functions of external field $E_e$, for three different values of $L=40$, 120 and 200~mm. We omit the asterisk in the notation of the optimal value (i.e. write $a$ instead of $a^\ast$ etc.). The dots highlight the same values as in Figures~\ref{fig:a_results_v}--\ref{fig:a_results}: (a)~streamer velocity $V$, (b)~the optimal radius $a$, (c)~the intrinsic field $E_s$, (d)~the maximum field $E_m$, (e)~the streamer electron density $n_s$, and (f)~the front thickness $d$ as a fraction of $a$ in percent. The measurements of \citet[eq. (6) for zero air humidity]{Allen+Mikropoulos1999} at $L=120$~mm are presented together with calculated $V$ results in panel (a).}
\label{fig:pos_results}
\end{figure*}

The results for positive streamers are presented in Figure~\ref{fig:pos_results}.  The velocity is compared to the measurement of \citet{Allen+Mikropoulos1999} who fitted the measured $V$ with a $E_e^3$ dependence. We observe that an exponential fit could also be valid. We must mention that \citet{Allen+Mikropoulos1999} obtained a velocity which is approximately constant with distance $L$, but our calculations suggest that it must grow with $L$. This may be explained by non-uniformity of the electric field in the experiment, since a pointed electrode was used to launch the streamer (although \citet{Allen+Mikropoulos1999} assert that the non-uniformity was confined to the first 1.5~cm of streamer propagation). Unfortunately, most of the streamer observations are done either in the lab with very non-uniform field created by electrodes shaped as a point \citep{Briels+2008}, wire \citep{Huiskamp+2017} or hemisphere \citep{Chen+2013}, or in the upper atmosphere in the form of sprites that propagate in a non-uniform medium, i.e., have lengths of the order of atmospheric scale height of $\sim 7$~km \citep{Kanmae+2012}, so we cannot compare these experimental results with our calculations at the present time. These comparisons are a subject of future work.

The inter-model verification (comparison with results of other numerical models) was beyond the scope of the present work, and is a subject of future research. Preliminary comparison with \citet{Bagheri+2018} shows $\sim$30\% difference in radius and velocity.

The maximum field at streamer head $E_m$ does not exhibit a lot of variation at $E_e\gtrsim 0.5$~MV/m. The constancy of $E_m$ gave rise to the ``chosen $E_m$'' hypothesis of \citet[p.~277]{book:Bazelyan+Raizer1998} which we criticize in this paper (see Subsection~\ref{ssec:other_authors}). The calculated typical value of streamer electron density $n_s\sim\sci{5}{18}$~m$^{-3}$ is of the same order as given by \citet[p.\ 343]{book:Raizer1991}.

\subsection{Negative streamers}
\label{ssec:neg_results}
\begin{figure*}
\includegraphics[width=0.98\textwidth]{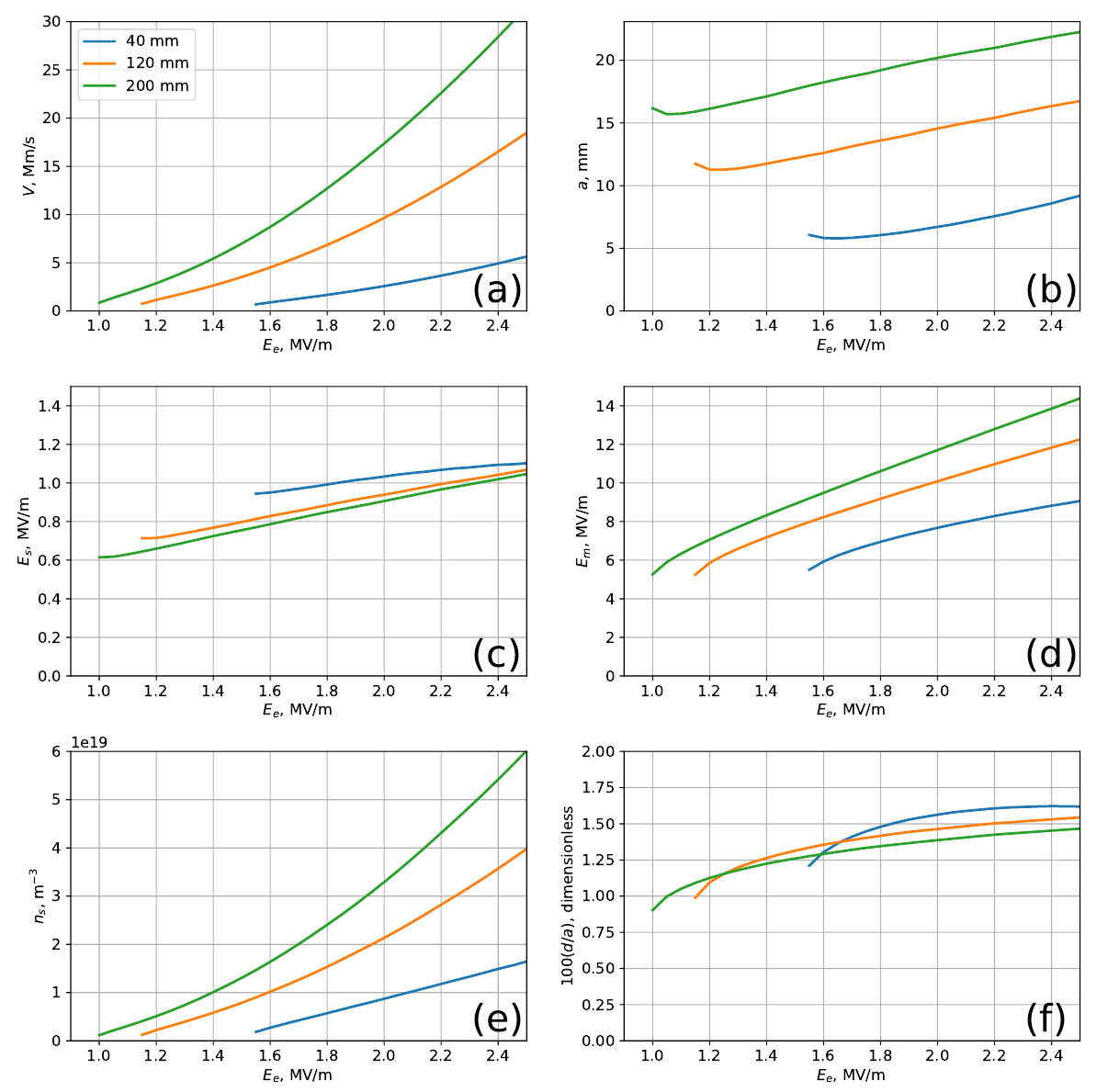}
\caption{Results for negative streamers (at maximum of $V$) as functions of external field $E_e$, for three different values of $L=40$, 120 and 200~mm. We omit the asterisk in the notation of the optimal value (i.e. write $a$ instead of $a^\ast$ etc.): (a)~streamer velocity $V$, (b)~the optimal radius $a$, (c)~the intrinsic field $E_s$, (d)~the maximum field $E_m$, (e)~the streamer electron density $n_s$, and (f)~the front thickness $d$ as a fraction of $a$ in percent.}
\label{fig:neg_results}
\end{figure*}

When we calculate streamer parameters as functions of $a$ for negative streamers, we also get, analogously to the positive streamer results, monotonous functions of $a$ for all parameters except $V$ which has a maximum at a certain value $a^\ast$. There are usually two branches, as depicted in Figure~\ref{fig:branches}, so we have to take $V^\ast$ on the upper branch. The results for negative streamers at $a^\ast$ are presented in Figure~\ref{fig:neg_results}. We omit the asterisk in the notation of the optimal value (i.e. write $a$ instead of $a^\ast$ etc.). Some of the quantities vary in a manner which is different qualitatively from the positive streamers: (i)~negative streamers of fixed length exist only above a certain threshold field $E_e$ (there is more on thresholds in Subsection~\ref{ssec:att}); (ii)~velocity dependence on $E_e$ is no longer well-fitted by an exponential, but is closer to linear; (iii)~the streamer radius $a^\ast$ starts with large values at low fields, unlike positive streamers in Figure~\ref{fig:pos_results}b; (iv)~the intrinsic field $E_s$ grows slower with $E_e$ than in the positive case; (v)~in contrast with the positive streamers, the maximum field at streamer head $E_m$ grows significantly with $E_e$, which may be used to refute the ``chosen $E_m$'' hypothesis of \citet[p.~277]{book:Bazelyan+Raizer1998}; (vi)~the calculated streamer electron density $n_s$ also grows with $E_e$, in contrast to the positive streamer result.

\section{Discussion}

\subsection{Discussion of the streamer radius results}
 \citet[p.~357]{book:Raizer1991} admits that the streamer radius is not very well known. The typical radius during the avalanche-to-streamer transition is given to be $a_0=0.18$~mm \citep[p.~332]{book:Raizer1991}. Hydrodynamic simulations \citep{Lehtinen+Ostgaard2018} indicate that it can be even smaller, a few tens of $\mu$m. The typical streamer radius at a more advanced stage is given in the range from 10--30~$\mu$m for short gaps \citep{Gallimberti+2002} to 0.5~mm \citep[p.~343]{book:Raizer1991}. Recent measurements suggest that streamers may be wider than given by these figures. E.g., \citet{Yi+Williams2002} have observed $a=3$~mm for positive and $a=4$~mm for negative streamers in 10\% O$_2$ mixture with N$_2$. \citet{Briels+2008} observed positive streamer diameters of 0.2--3~mm in a needle-plane electrode geometry in a 40~mm gap.  \citet{Chen+2013} observed positive streamers with diameters 1.6--6.3~mm for a 2-cm-diameter hemispherical electrode 2~cm and streamer length of 2--16~cm.

Very wide diameters, $\gtrsim 1$~cm were measured by \citet{Tarasenko+2018}, for both positive and negative streamers. The negative streamers have larger diameters for the same conditions (geometry and voltage), as seen from comparison of their Figures~4 and~5. The measured diameter and velocity increase as a function of applied voltage and wider streamers move faster than thinner streamers, both in the sprite streamers and in laboratory discharges \citep{Kanmae+2012}. The photographic observations of \citet{Kochkin+2016} indicate radii up to $\sim$1~cm.

The streamer radius calculated here is rather large ($\gtrsim$1~cm) for negative streamers and for some values of $E_e$ for positive streamers. The dependence on field and correlation with velocity in our results presented in Figure~\ref{fig:pos_results}a and~\ref{fig:pos_results}b agree with the above-mentioned experimental trends \citep{Kanmae+2012}. Radii for negative streamers are obtained to be larger than those of positive streamers (compare Figures~\ref{fig:neg_results}b and~\ref{fig:pos_results}b), in agreement with experiment \citep{Tarasenko+2018}.

\subsection{Electron attachment in the channel and streamer threshold fields}
\label{ssec:att}
The electrons inside the streamer channel will eventually attach. (For estimates in this Subsection, we disregard the processes of detachment and recombination.) The length over which it occurs is
\begin{equation}
  L_\attach = \frac{V\pm v(E_s)}{\nu_a(E_s)-\nu_i(E_s)}
  \label{eq:Latt}
\end{equation}
This expression is valid only when it is $>0$ (i.e., $\nu_a>\nu_i$ in the channel). We took into account both the effects of the streamer motion and the electron drift.

\begin{figure*}
\includegraphics[width=0.98\textwidth]{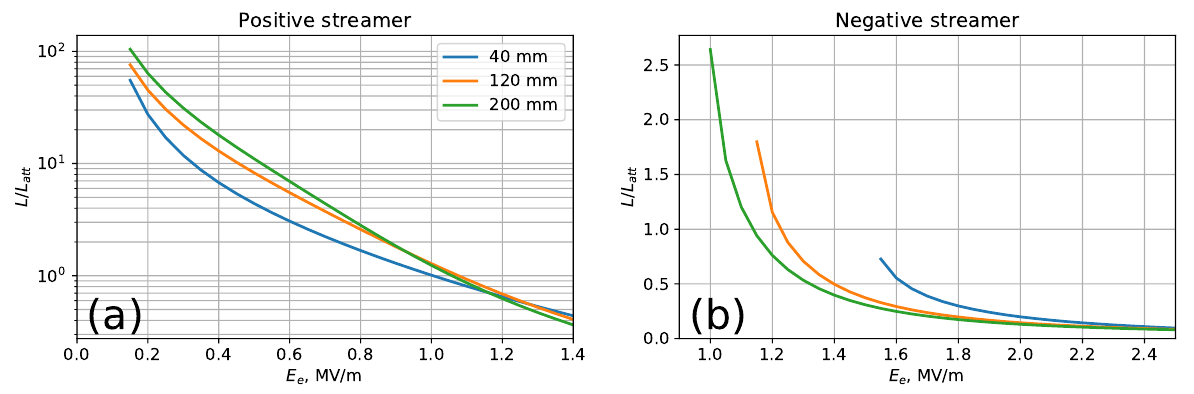}
\caption{The number of attachment lengths in (a)~positive and (b)~negative streamers.}
\label{fig:natt}
\end{figure*}

\begin{figure}
\includegraphics[width=0.49\textwidth]{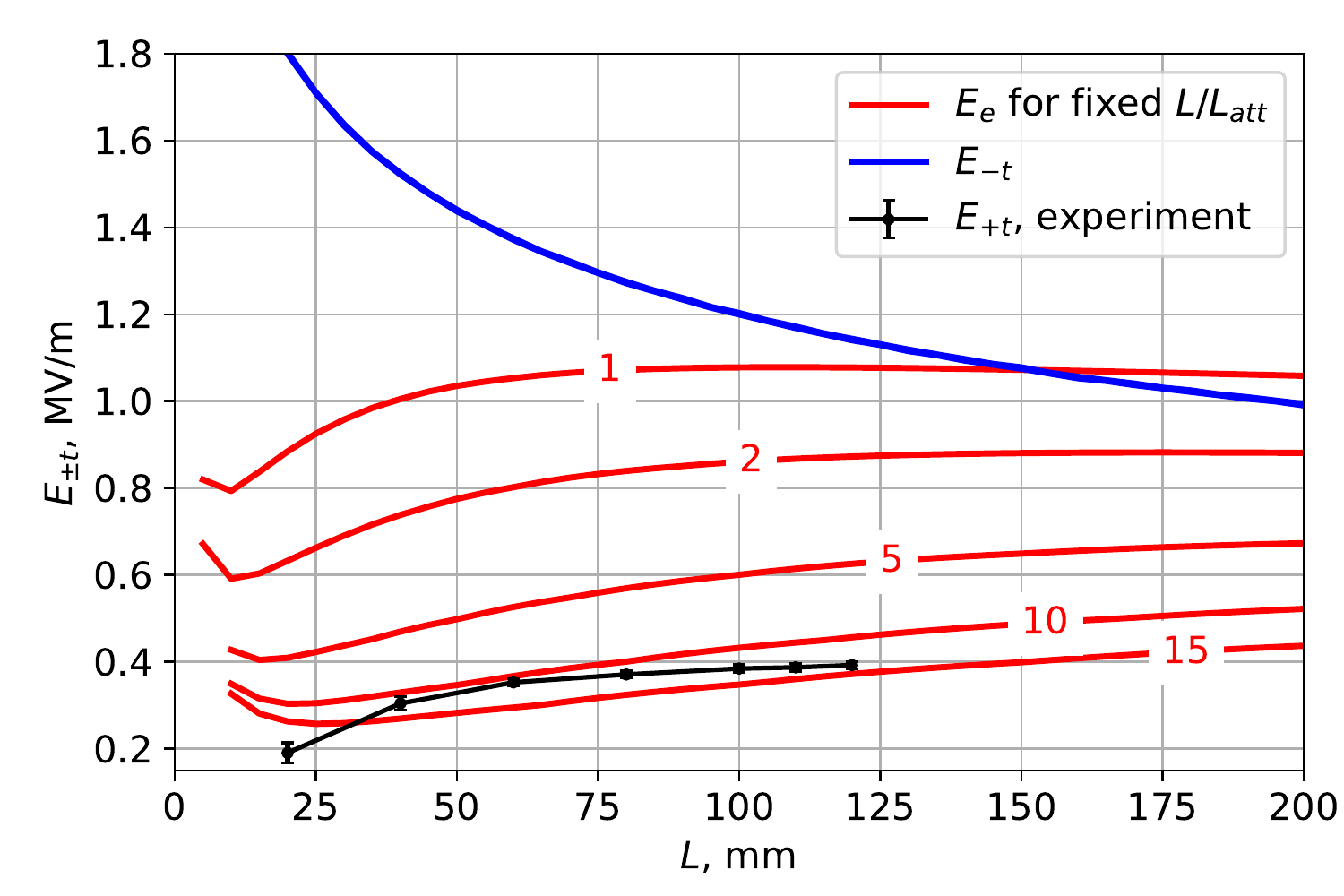}
\caption{Positive and negative streamer threshold fields $E_{\pm t}$, as a function of the streamer length $L$. The experimental data are from \citet[Fig.~4]{Allen+Mikropoulos1999}.}
\label{fig:Ethresh}
\end{figure}

The calculated number of attachment lengths $L/L_\attach$ is plotted in Figure~\ref{fig:natt}, for both positive and negative streamers.  The conductivity in the streamer channel drops $\propto e^{-L/L_\attach}$, and when it becomes too low to support the streamer current, this should lead to electric detachment of the streamer from the electrode and probably to the eventual quenching of the streamer. This may explain the existence of threshold $E_e$ for the positive streamers. Thus, we may estimate the positive streamer threshold field $E_{+t}$ by finding $E_e$ at which $L/L_\attach$ achieves a fixed value close to 1. Such curves are plotted in Figure~\ref{fig:Ethresh}. To compare, we also plot the experimentally measured threshold fields \citep[Fig.~4, 270~ns triggering pulse]{Allen+Mikropoulos1999}. The error bars are such that lower and upper limit correspond to what was called ``threshold'' and ``stability'' fields by \citet{Allen+Mikropoulos1999}. \citet{Aleksandrov+Bazelyan1999} suggested that electron-ion recombination is more important for the streamer channel decay than attachment, which may explain why the value of $L/L_\attach\sim 1$ does not match well the experimental data, but instead the best-matching values are $\sim 10$--15. Anyway, both are three-body processes, so we cannot expect $E_{+t}$ to be simply proportional to atmospheric density, but instead to have a nonlinear dependence \citep{Phelps+Griffiths1976, Aleksandrov+Bazelyan1996}.

We must point out that quenching of the streamer when it becomes electrically detached from the electrode is not proven yet. In fact, \citet[Figure~22b]{Gallimberti1979} describes a unipolar streamer soliton with a ``compensation zone'' behind the streamer in which the conductivity and charge gradually fall to zero, which contradicts our statement that such objects must be quenched. However, we were not able to find any numerical confirmation in literature of existence of such solitons. Of course, bipolar electrode-free streamer systems (pilots) do exist \citep{Kochkin+2016,Lehtinen+Ostgaard2018} and are an important stage in propagation of negative leaders \citep{Gorin+1976,Gallimberti+2002}.

For negative streamers, $L/L_\attach$ stays mostly $\lesssim 1$, as seen in Figure~\ref{fig:natt}b, and the threshold is determined by a different mechanism. Namely, as we see in Figure~\ref{fig:neg_results}, the solution is disappearing at small $E_e$. This happens because the two negative streamer mode branches in Figure~\ref{fig:branches}, become closer and closer to each other as $E_e$ is lowered and eventually disappear at $E_e=E_{-t}$, which we define to be the negative streamer threshold field. The value of $E_{-t}$ calculated in this way is also plotted in Figure~\ref{fig:Ethresh}. The results are compatible with the commonly accepted values of the the negative streamer threshold fields $E_{-t}=0.75$--1.25~MV/m \citep[p.~362]{book:Raizer1991}. Interestingly, $E_{-t}$ decreases with $L$, which is the opposite behavior from $E_{+t}$. This may mean that in order for a negative streamer to grow starting from zero length, the external field with average values in the range of Figure~\ref{fig:Ethresh} ($E_e\lesssim 2$~MV/m) must be non-uniform (stronger closer to the electrode).

\subsection{Streamer branching}
The model proposed in this paper may be capable of describing the branching of streamers, even though we represented a streamer as a single conducting cylinder. Branching is an instability when the former streamer radius is suddenly replaced by a new, smaller one, so that the streamer necessarily has to split into two (or more) branches. Such an instability could have occurred in the presented theory, if the dispersion function $V(a)$ had a shape with two peaks, with their relative height changing as a function of $L$. If at smaller $L$ the peak at larger $a$ was higher, but above some critical $L$ the peak at the smaller $a$ became higher, then it would mean that the optimal $a^\ast$ would change suddenly (discontinuously) from a higher to lower value. If the lower value $a^\ast$ were a fraction of the higher one, this could lead to splitting of the channel.

However, this condition never occurred in the presented results: the dispersion function $V(a)$ always had a single peak (e.g., see Figures~\ref{fig:branches},~\ref{fig:a_results_v}). We may hypothesize that branching has a difficulty to occur in a constant uniform field. The observations of branching usually involve electrodes of limited size and changing voltage which implies existence of a non-uniform and non-constant electric field. More uniform field configurations generally lead to less branching \citep{vanVeldhuizen+Rutgers2002}. Another possibility is a non-uniformity in medium. Such a situation can occur, e.g., in sprites, giant discharges occurring at mesospheric altitudes which span several atmosphere scale heights \citep{Pasko+2013}.  It is, however, also possible that some mechanism is responsible for a local instability at the streamer head when the head becomes too wide, and that was not taken into account in the presented model. At the present time, no real physical understanding of streamer branching exists, and the results of \citet{Devauchelle+2012} for a two-dimensional diffusion-limited aggregation (DLA) system seem to be not applicable. In particular, \citet[p.~292]{book:Pasko2006} reported that ``the results on branching morphology reported by different research groups remain highly controversial'' and may be due to numerical effects.

\section{Summary and future work}

\subsection{Summary}
We have identified a system of algebraic equations (SAE) (\ref{eq:system}) relating streamer parameters to each other, which follows from hydrodynamic description of the streamer discharge attached to an ideal electrode and propagating in constant uniform external field. This was done by assuming an ionization front in a shape of a column, and by making a row of approximations. The approximation which gives the biggest error (of a few tens of percent) is most likely that of uniformity of electron density along the streamer channel, as we neglected its time evolution due to streamer propagation (Subsubsection~\ref{sssec:Es_ns_uniformity}) and processes of electron attachment and recombination in the channel (Subsection~\ref{ssec:att}). The error may be acceptable for practical application of the method, because the alternative, namely numerical solutions of hydrodynamic equations, may also show considerable variations ($\sim 10$\% inferred from figures in \citep{Bagheri+2018}).

The streamer modes may be found by solving this system for all streamer parameters, with given the external field $E_e$ (determined by laboratory conditions), streamer length $L$ (determined by its history of propagation) and streamer radius $a$. The preferred mode is determined by the velocity maximization (max-$V$) criterion.  The roles of various physical processes are made transparent by writing SAE in a form which may be solved analytically (Appendix~\ref{app:analytic}), even though this increases the error.

The negative streamer threshold field is defined by disappearance of solutions (modes) at low $E_e$ and $L$.

Our model does not predict branching in the considered conditions (uniform constant $E_e$ and ideal electrode), but, instead, stable propagation of a single mode at all $L$ because $V(a)$ has a single local maximum.

\subsection{The place of the presented model in the roadmap of theoretical streamer studies}
Most modern theoretical studies of streamers that aim at reproducing the correct physical picture are mostly done by hydrodynamic numerical simulations, which are at the verge of quantitative validation by comparison with experiment but still may be not sufficiently accurate \citep{Bagheri+2018}. These codes calculate various physical values (such as field, electron density etc.) at a finite number of grid points, thus storing and updating (as a function of time) a finite set of variables. Stability and accuracy requires high density of grid points to resolve sharp gradients, and therefore a large number of variables. For this reason, such numeric codes need extensive computational resources to simulate even a single streamer. Increasing accuracy of represented physics by considering kinetic equations requires either solving more complicated partial differential equations, or following individual particles (as in particle-in-cell approaches), and therefore even more dynamic variables and more computational resources.

However, if we expect a solution of a fixed type, namely that of an ionization front with a shape of a column, many of these variables are strongly correlated with each other (e.g., the electron density inside the channel changes slowly from point to point). All required information therefore may fit into a smaller number of variables. This is exactly what we suggest in this paper, by describing a streamer with a very limited set of streamer parameters (such as velocity, radius, etc.).

To increase accuracy, more variables may be added to the description. For example, the biggest error, associated with the change of electron density along the channel, may be remedied by making it into a one-dimensional dynamic variable. If the presented relations may be modified to take into account non-uniform external fields, it would become possible to describe more complicated configurations than just a single streamer. This would require limited computational resources, similar to other one-dimensional approaches \citep{Aleksandrov+Bazelyan2000, Samusenko+Stishkov2011, Luque+Ebert2014}, but, unlike them, rooted only in physics instead of relying on external parametrization of some unknowns. Thus, we are convinced that, despite having limited accuracy, our model has good perspectives in the roadmap of theoretical streamer studies.

\appendix
\section{Ionization front with nonzero current}
\label{app:ns}
Let us estimate the field and electron density on the axis of the streamer, while taking into account that there is a total current flowing along the axis, given by equation (\ref{eq:J}). We emphasize that although equations are solved in 1D, the system is not translationally symmetric in the transverse direction as the radius of the streamer is finite.

\subsection{General solution}
With nonzero total current $J$, instead of equations (\ref{eq:front}), from (\ref{eq:hydro_front}) we get
\begin{equation}
  \left.\begin{array}{rcl}
  \varepsilon_0 V\partial_\xi E &=& e n v  - J \\
  \partial_\xi \left[(V\pm v) n\right] &=& -\alpha_t v n
  \end{array}\right\}
  \label{eq:app:front_with_current}
\end{equation}
where $\alpha_t(E)$ is the net spatial ionization rate and $J$ is given by equation (\ref{eq:J}). When taking values on the axis, we neglected the transverse contribution to the divergence in the second equation. Let us multiply the first equation by $\alpha_t/(e V)$ and add it to the second equation multiplied by $1/V$. Utilizing the ionization integral (\ref{eq:Psi}) for which $\partial_\xi\Psi = \alpha_t\partial_\xi E$, we get
\[ \partial_\xi\left[\left(1\pm\frac{v}{V}\right)\frac{en}{\varepsilon_0} + \Psi(E)\right] = -\frac{J\alpha_t(E)}{\varepsilon_0 V} \]
Integrating and fixing the integration constant so that $n=0$ at $\xi=+\infty$ (no ionization far ahead, also $E=0$ there), we have:
\begin{equation}
  \frac{e n(E,\xi)}{\varepsilon_0} = \left(1\pm\frac{v}{V}\right)^{-1} \left[\int_\xi^\infty \frac{J(\xi')\alpha_t(E)\,d\xi'}{\varepsilon_0 V} - \Psi(E)\right]
  \label{eq:n_J}
\end{equation}
This may be substituted into the first equation in the system (\ref{eq:app:front_with_current}) which now contains only a single unknown $E$. However, it can be solved only numerically, which was outside the scope of this paper and may be a part of future effort to make the presented theory more accurate. We still can get some insight into how $J$ affects the relation between $n_s$ and $E_f$ given by equation (\ref{eq:ns}) by analyzing the system (\ref{eq:app:front_with_current}) with some approximations.

\subsection{Solution behind the front}
At this point, let us neglect $v\ll V$ in the second equation of system (\ref{eq:app:front_with_current}) and therefore in the denominator of (\ref{eq:n_J}). Then, substituting expressions (\ref{eq:Psi}) for $\Psi$ and (\ref{eq:J}) for $J$, we get
\begin{equation}
  n(E,\xi) = \frac{\varepsilon_0}{e}\left[\int_\xi^\infty \frac{(E_f-E_e)\alpha_t\,d(\xi'/l)}{(1+\max\{0,\xi'\}/l)^2} - \int_0^E \alpha_t\,d E\right]
  \label{eq:app:n_J_approx}
\end{equation}
This is still a very complicated, but we may make some rough estimates if we assume a simple dependence $\alpha_t(E)$. For example, assume threshold behavior $\alpha_t(E)=\alpha=\const>0$ at higher fields $E>E_1$ and $\alpha_t(E)=0$ at lower fields $E<E_1$. Since $E$ decreases within the front towards the tail, we can take $\alpha_t(E)=\alpha$ at $\xi>\xi_1$ and $\alpha_t=0$ at $\xi<\xi_1$, where $E(\xi_1)=E_1$ and $\xi_1$ is somewhere within the ionization front, i.e., $\abs{\xi_1}\lesssim d$, the thickness given by equation (\ref{eq:d}). The streamer electron density $n_s=n(\xi\rightarrow-\infty)$ may then be calculated from (\ref{eq:app:n_J_approx}):
\[
  n_s \approx  \frac{\varepsilon_0}{e}(E_f-E_e)\alpha\times\left\{\begin{array}{lr}
    1+\abs{\xi_1}/l, & -d\lesssim\xi_1<0 \\
    1/(1+\xi_1/l), & 0<\xi_1\lesssim d \end{array}\right.
\]
This must be compared with an expression that we would get for a flat front:
\[ n_s^\mathrm{flat}=\frac{\varepsilon_0}{e}(E_f-E_1)\alpha \]
Within an error of the order of
\begin{equation}
  \frac{\abs{n_s^\mathrm{flat}-n_s}}{n_s^\mathrm{flat}}\lesssim\max\left\{\frac{\abs{E_e-E_1}}{E_f},\frac{d}{l}\right\}\ll 1
  \label{eq:app:n_J_error}
\end{equation}
the two expressions are the same.

\subsection{Solution ahead of the front}
In this Subsection, let us neglect $E_1,E_e\ll E_f$ and consider (\ref{eq:app:n_J_approx}) only for $\xi>\max\{0,\xi_1\}$, where we have $\alpha=\const$. Again, the integration may be performed and we have
\begin{equation}
  \frac{n}{n_s} =  \frac{1}{1+\xi/l} - \frac{E}{E_f},\quad n_s=\frac{\varepsilon_0\alpha E_f}{e}
  \label{eq:n_J_ahead}
\end{equation}
Before we proceed, it is instructive to convert the variables in (\ref{eq:app:front_with_current}) to dimensionless. We introduce
\[ x=\xi/d,\quad y=E/E_f,\ z=n/n_s \]
where $d$ is given by equation (\ref{eq:d}):
\[ d=\frac{V \pm v(E_f)}{\alpha v(E_f)}\approx \frac{V}{\alpha v(E_f)} \]
There is a small parameter
\[ \delta\equiv \frac{d}{l} \approx \frac{v(E_s)}{v(E_f)} \ll 1\]
where the second expression was obtained by using $J_0 = e n_s v(E_s)\approx\varepsilon_0 V E_f/l$ (see equation \ref{eq:J0}) and the above expression for $n_s$. Equation (\ref{eq:n_J_ahead}) becomes
\[ z = \frac{1}{1+\delta x}-y \]
and the first equation of (\ref{eq:app:front_with_current}) becomes after substitution of $n(E)$
\begin{equation}
  \frac{d y}{d x} = y z - \frac{\delta}{(1+\delta x)^2} = y\left(\frac{1}{1+\delta x}-y\right) -  \frac{\delta}{(1+\delta x)^2}
  \label{eq:y}
\end{equation}
Let us switch to variable $z$. Then
\[ \frac{d z}{d x} = -y z = -z\left(\frac{1}{1+\delta x}-z\right) \]
This is a Bernoulli equation which may be solved by substituting $u=1/z$, for which we get a linear inhomogeneous first-order equation
\[ \frac{d u}{d x} - \frac{u}{1+\delta x} = -1 \]
For $\delta=0$ (which corresponds to the flat front without currents), the solution is in the shape of the sigmoid curve:
\begin{equation}
  z(x) = \frac{1}{1+e^{x-x_0}},\quad y(x) = \frac{1}{1+e^{-(x-x_0)}},\quad x_0=0
  \label{eq:sigmoid}
\end{equation}
where $x_0$ is the constant of integration which we fixed on the consideration that $x=0$ represents the front.

For $\delta>0$, the solution is
\[ z(x) = (1+\delta x)^{-1}\left[\frac{1}{1-\delta}+C(1+\delta x)^{1/\delta-1}\right]^{-1} \]
\[ y(x) = (1+\delta x)^{-1}\frac{\frac{\delta}{1-\delta}+C(1+\delta x)^{1/\delta-1}}
	{\frac{1}{1-\delta}+C(1+\delta x)^{1/\delta-1}} \]
The solution for $y(x)$ has a shape (\ref{eq:E}) where $(1+\delta x)^{-1}$ represents the $\propto(1+\xi/l)^{-1}$ part while the rest is the switch function $S(\xi)$. The constant $C$ should be fixed from the consideration that $x=0$ represents the front, and should be the same as (\ref{eq:sigmoid}) in the limit $\delta\rightarrow 0$. Our goal is to find the location where $y(x)$ is maximized. We will show that it happens at $1\ll x\ll 1/\delta$, so if we make an error in the position of the front $\Delta x_0\lesssim 1$, it will not affect our result significantly. This gives us certain freedom in the choice of $C$. For example, we may choose $C=1/(1-\delta)$:
\begin{equation}
  y(x) = \frac{1}{1+\delta x}\frac{\delta+(1+\delta x)^{1/\delta-1}}{1+(1+\delta x)^{1/\delta-1}}
  \label{eq:y_ahead}
\end{equation}
Neglecting higher orders in $\delta$, we find that the maximum of this function is located at $1+\delta x_m \approx e^{\delta\log(1/\delta)}$. Expanding the exponent in Taylor series we get $x_m\approx \log(1/\delta)$ and hence the maximum value of $y_m = y(x_m) \approx e^{-\delta\log(1/\delta)} \approx 1 - \delta\log(1/\delta)$. Going back to dimensional variables, we obtain that the maximum field $E_m$ is attained at $\xi=\xi_m$, where
\begin{equation}
  \frac{E_m}{E_f}=1-\frac{d}{l}\log\left(\frac{l}{d}\right),\mbox{ at }\xi_m = d\log\left(\frac{l}{d}\right)
  \label{eq:app:Em}
\end{equation}
Since we neglected $E_e\ll E_f$, this equation also could have $(E_m-E_e)/(E_f-E_e)$ on the left-hand side. The results presented in Figures~\ref{fig:pos_results}d, \ref{fig:neg_results}d do not change significantly if we make this substitution.

Using a well-known identity $\lim_{\delta\rightarrow 0}(1+\delta x)^{1/\delta}= e^x$, (\ref{eq:y_ahead}) for small $\delta$ gives approximately
\[ y(x) \approx \frac{1}{(1+[1+\delta x]e^{-x})(1+\delta x)} \]
This may be considered as the shape of $E(\xi)$ given by equation (\ref{eq:E}), with the switch function given by
\begin{equation}
  S(\xi)=\left[1+(1+\xi/l)e^{-\xi/d}\right]^{-1},\quad \xi\gtrsim \max\{0,\xi_1\}
  \label{eq:app:S}
\end{equation}
The value of $\xi_1$ is within $\abs{\xi_1}\lesssim d$, so we may use this expression for $\xi\gtrsim 0$.

\subsection{Summary}
We analyzed an ionization front with current (\ref{eq:J}) flowing through it. The maximum electric field is attained slightly ahead of it at position $\xi_m$ and has value $E_m$, which is smaller than $E_f$ and is given by equation (\ref{eq:app:Em}).  The electron density behind the front $n_s$ is still determined by $E_f$ and is given by equation (\ref{eq:ns}), with error given by (\ref{eq:app:n_J_error}). The relative change in $n_s$ due to the current is much smaller than the relative deviation of $E_m$ from $E_f$.

\section{Streamer velocity in analytical approximation}
\label{app:analytic}
Here we demonstrate mostly qualitatively the main results of the paper, namely, that the dispersion function $V(a; E_e, L)$ has a maximum and the max-$V$ velocity $V^\ast$ and radius $a^\ast$ have qualitative behavior as functions of $E_e$ and $L$ that is consistent with the more accurate results presented in the main body of the paper (and experiments). We also hope that this provides additional clarity and support for soundness of claims made in this paper.

We make several simplifying assumptions, in order to keep our equations tractable:
\begin{enumerate}
\item Mobility $\mu=\const\approx 0.05$~m$^2$\,V$^{-1}$\,s$^{-1}$.
\item Neglect attachment, and let us denote $\alpha_f=\alpha_t(E_f)$ for brevity (we remind that $\alpha_t$ is the first Townsend coefficient, i.e., net spatial ionization rate). In some formulas that we will need, $\alpha_t$ is averaged over $E$ in interval $[0,E_f]$, weighted with a power of $E$. Let us introduce a dimensionless coefficient $K_m$, which is a function of $E_f$:
\begin{equation}
  K_m \equiv \frac{1}{\alpha_f E_f^m}\int_0^{E_f}\alpha_t(E)E^{m-1}\,d E
  \label{eq:app:K}
\end{equation}
\item Method-of-moments results:
\begin{itemize}
\item $\eta\approx L/a$, which is an approximate form of equation (\ref{eq:eta}). Reasoning for why it has this approximate form may be found in \citep[p.~78]{book:Bazelyan+Raizer1998}.
\item $\chi\equiv l/a \approx 1/\sqrt{8}=\const$ (equation \ref{eq:l_approx}).
\end{itemize}
\item $E_s,E_e\ll E_f$.
\end{enumerate}

We use approximate versions of equations (\ref{eq:system}, \ref{eq:system_E}):
\begin{enumerate}
\item Equation (\ref{eq:system_E}) for $E$ near streamer tip is approximately:
\begin{equation}
  E(\xi) \approx \frac{E_f}{1+\xi/l}
  \label{eq:app:E}
\end{equation}
where we neglected all terms $\sim 1$ compared to $\sim\eta$, and the approximate form of equation (\ref{eq:Ef}) is
\begin{equation}
  E_f \approx \eta(1-\zeta)E_e
  \label{eq:app:Ef}
\end{equation}
where we introduced $\zeta\equiv E_s/E_e<1$.
\item Equation of current continuity (\ref{eq:Jcontinuity}):
\begin{equation}
  \frac{e n_s}{\varepsilon_0} \mu \zeta = \frac{V}{l}\eta(1-\zeta)
  \label{eq:app:Jcontinuity}
\end{equation}
\item Equation (\ref{eq:ns}) becomes, using (\ref{eq:app:K}): 
\begin{equation}
  \frac{e n_s}{\varepsilon_0} \approx \int_0^{E_f} \alpha(E)\,d E = K_1 \alpha_f E_f
  \label{eq:app:ns}
\end{equation}
\item \citet{Loeb1965} equation (\ref{eq:Loeb65}) replaces \citet{Pancheshnyi+2001} equation (\ref{eq:photobalance}). We keep the integral:
\[ V = \frac{1}{\log(n_s/n_p)}\int_0^{\xi_p}\nu_t(\xi)\,d\xi \]
We can use $p\equiv \log(n_s/n_p)\approx 8$ given by \citet{Naidis2009}, which may be derived (approximately) by substitution $\xi_p=\aph$ in equation (\ref{eq:nphoto}) (this value is suggested by results of Subsection~\ref{ssec:photodistance}). This derivation is not included here.

From (\ref{eq:app:E}), in the same way as \citet{Naidis2009}, we get:
\[ d\xi = -\frac{l E_f d E}{E^2}\]
substituting which, and also $\nu_t(E) = \mu E \alpha_t(E)$:
\[ \int_0^{\xi_p}\nu_t(E)\,d\xi=\mu E_f l\int_{E(\xi_p)}^{E_f} \alpha_t(E)\,\frac{d E}{E} \approx K_0\nu_f l  \]
where we neglected the field at $\xi_p$ and introduced the maximum ionization rate at the streamer front
\[ \nu_f=\nu_t(E_f)=\mu E_f\alpha_f \]
\citet{Loeb1965} equation becomes, finally
\begin{equation}
 V = \frac{K_0 l\nu_f}{p}
 \label{eq:app:photobalance}
\end{equation}

\end{enumerate}

To reduce everything to a single equation, substitute (\ref{eq:app:ns}) and (\ref{eq:app:photobalance}) into (\ref{eq:app:Jcontinuity}), which we can solve for $\zeta$:
\[ \zeta = \frac{E_s}{E_e} = \frac{1}{1+q/\eta},\quad q\equiv\frac{K_1p}{K_0} \]
This is not a simple expression of $E_s$ as a function of $a$ because $K_m$ depends on unknown $E_f$. However, we may guess approximate values of $K_m$ to be used. Using Townsend formula $\alpha_t(E)\approx\alpha_{i0} e^{-E_{i0}/E}$ with $\alpha_{i0}=\sci{5.4}{5}$~m$^{-1}$, $E_{i0}=19.5$~MV/m (see equation \ref{eq:nu_i}), we get
\[ K_m = \frac{1}{y^m}\int_0^y x^{m-1}e^{-1/x}\,d x,\quad y=\frac{E_f}{E_{i0}}\approx 0.6\mp 0.2 \]
where the approximate range of $y$ is taken from results in this paper (or other works). Numerically, we get
\[ K_0\approx 0.4\mp 0.1,\quad\frac{K_1}{K_0} \approx 0.75\pm 0.05,\quad q \equiv \frac{K_1p}{K_0} \approx 6  \]
We take advantage of $q$ varying rather slowly with $E_f$. Thus, we have found (after also substituting $\eta=L/a$):
\begin{equation}
  E_s(a) = \frac{E_e}{1+q a/L},\quad q\approx 6
  \label{eq:app:Es}
\end{equation}
We are interested in finding and maximizing $V(a)$. To find it, we first need, obtained from (\ref{eq:app:Ef}):
\[ E_f(a) = \frac{q E_e}{1+q a/L} \]
Velocity is expressed from (\ref{eq:app:photobalance}):
\[ V(a) = \left[\frac{\chi\mu\alpha_{i0}K_0}{p}\right] a E_f(a)e^{-E_{i0}/E_f(a)} \]
We use the fact that $K_0$ does not vary much, so to find the maximum we may consider only the non-constant part that comes after the brackets, and take the term in brackets to be constant (with $K_0=0.4$). This function does, in fact, have a maximum. It was essential that ionization rate had Townsend shape. For $\alpha_t(E)\approx\const$, for example, we would not have a maximum in $V(a)$, which means that, potentially, there can exist systems in which streamers cannot form.

Returning to our task of finding the maximum of $V(a)$, we make another rough approximation $E_f\approx\const$ when it is the multiplier and write it as
\[ V(a) = \const\times a e^{-c(1+q a/L)},\quad c = \frac{E_{i0}}{q E_e} \]
This function has a maximum at
\begin{equation}
  a^\ast = \frac{L E_e}{E_{i0}}
  \label{eq:app:a}
\end{equation}
We immediately see the expected behavior $a^\ast\propto L, E_e$. Moreover, the streamer, as expected, is narrow ($a\ll L$) for usual experimental values of the electric field $E_e\ll E_{i0}$. Substituting this into the expression for $E_f$ we have
\[ E_f^\ast\equiv E_f(a^\ast) = \frac{q E_e}{1+q E_e/E_{i0}} \]
After substitution in expression for $V$:
\begin{equation}
  V^\ast\equiv V(a^\ast) = \left[\frac{\chi\mu\alpha_{i0}K_0}{p}\right] \frac{L E_e^2e^{-1-E_{i0}/(q E_e)}}{E_e + E_{i0}/q}
  \label{eq:app:V}
\end{equation}
We see the expected behavior $V^\ast\propto L$, which, by the way, is also characteristic for the small transverse perturbations of a flat front, see equation (\ref{eq:maxVflat}). Dependence on $E_e$ is a bit more complicated. For large $E_e$ ($E_e>E_{i0}/q\approx 3.3$~MV/m), we have approximately $V\propto E_e$, while for small $E_e$ ($E_e<3.3$~MV/m), it is $V\propto E_e^2e^{-E_{i0}/(q E_e)}$. We remind that in the experiment \citep{Allen+Mikropoulos1999}, the dependence $V\propto E_e^3$ for fields $0.4<E_e<0.8$~MV/m was observed.

\begin{figure}
\includegraphics[width=0.49\textwidth]{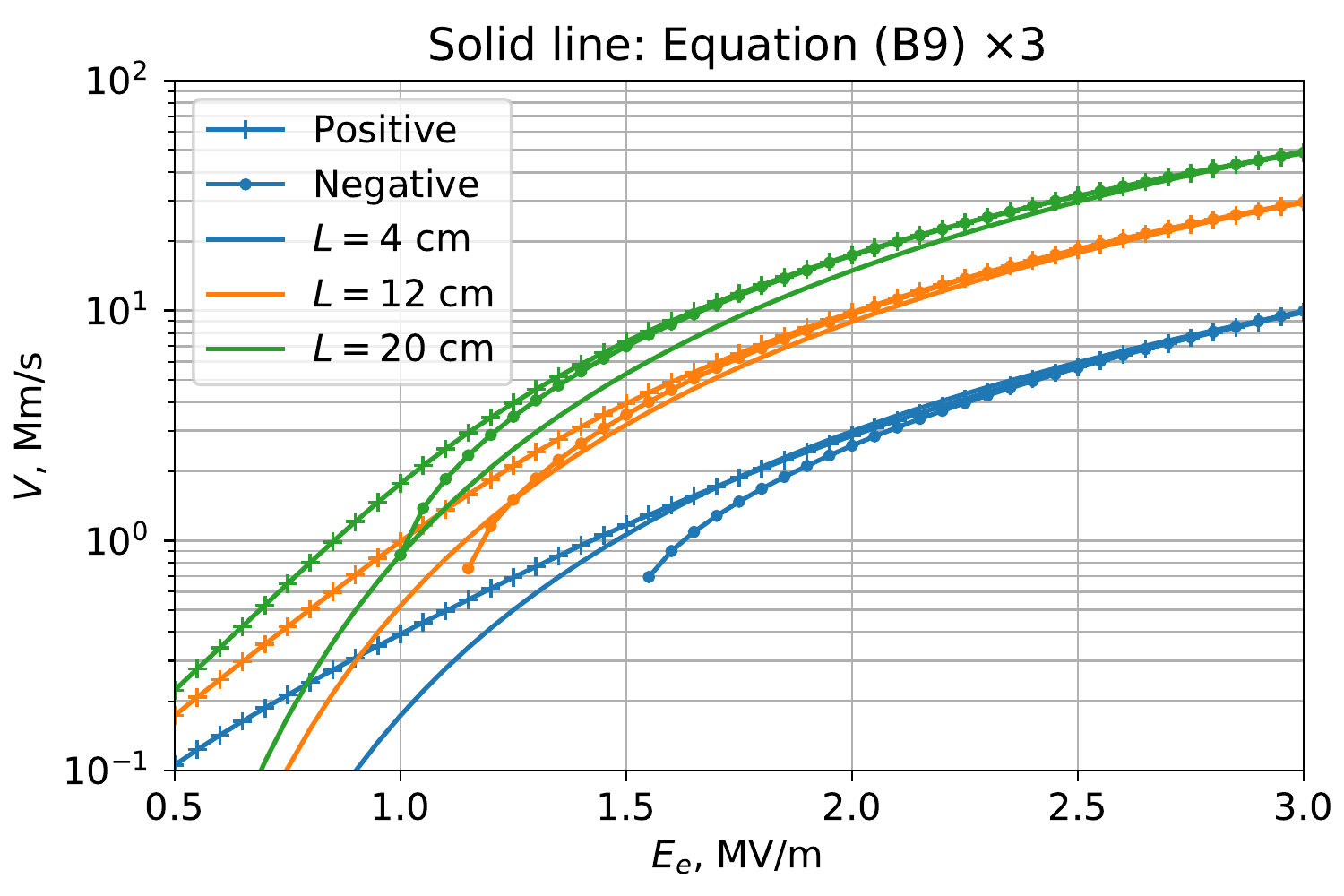}
\caption{Streamer velocity calculated with approximate formula (\ref{eq:app:V}), scaled by $\times$3, compared to the more accurate results of Figures~\ref{fig:pos_results} and~\ref{fig:neg_results}.}
\label{fig:appendix_b}
\end{figure}

Thus, we have reproduced, at least qualitatively, many of the results of this paper, only by using simple analytical formulas. Among the things that we did not reproduce are, e.g., the two branches of $V(a)$ for negative streamers and the existence of the negative streamer threshold field $E_{-t}$. This is mainly because we neglected electron drift velocity in equation (\ref{eq:app:photobalance}), so that we effectively consider a ``sign-less'' (as opposed to positive or negative) streamer. Also, because of many other approximations, when compared to the results in Figures~\ref{fig:pos_results} and \ref{fig:neg_results}, the velocity given by equation (\ref{eq:app:V}) is $\sim$3~times smaller in the limit of higher $E_e$, as demonstrated in Figure~\ref{fig:appendix_b}. At lower fields, $E_e\lesssim 1.5$~MV/m, the discrepancy is even higher. Unfortunately, if we remove any of the approximations, we will not be able to write a simple analytical formula like (\ref{eq:app:V}).

\section{Implementation}
\label{app:code}
The Python3 code implementing the method described in this paper is attached to the submission, and is also available at \url{https://gitlab.com/nleht/streamer_parameters}, with instructions on how to reproduce the presented results.

\begin{acknowledgments}
This study was supported by the European Research Council under the European Union's Seventh Framework Programme (FP7/2007-2013)/ERC grant agreement n.~320839 and the Research Council of Norway under contracts 208028/F50, 216872/F50 and 223252/F50 (CoE).
\end{acknowledgments}

\bibliography{../streamer_parameters}
\end{document}